\documentclass[twocolumn]{aastex631}
\usepackage{amsmath}

\begin{document}

\title{Illuminating the Incidence of Extraplanar Dust Using Ultraviolet Reflection Nebulae with \textit{GALEX}}

\correspondingauthor{Erin Boettcher}
\email{eboettch@umd.edu}

\author[0000-0003-3244-0409]{Erin Boettcher}
\affiliation{Department of Astronomy, University of Maryland, College Park, MD 20742, USA}
\affiliation{X-ray Astrophysics Laboratory, NASA/GSFC, Greenbelt, MD 20771, USA}
\affiliation{Center for Research and Exploration in Space Science and Technology, NASA/GSFC, Greenbelt, MD 20771, USA}

\author[0000-0002-2397-206X]{Edmund Hodges-Kluck}
\affiliation{X-ray Astrophysics Laboratory, NASA/GSFC, Greenbelt, MD 20771, USA}



\begin{abstract}

Circumgalactic dust grains trace the circulation of mass and metals between star-forming regions and gaseous galactic halos, giving insight into feedback and tidal stripping processes. We perform a search for ultraviolet (UV) reflection nebulae produced by extraplanar dust around 551 nearby ($D < 100$ Mpc), edge-on disk galaxies using archival near-UV (NUV) and far-UV (FUV) images from \textit{GALEX}, accounting for the point-spread function (FWHM $= 4''-5''$). We detect extraplanar emission ubiquitously in stacks of galaxies binned by morphology and star-formation rate, with scale heights of $h_{\text{h}} = 1 - 2.3$ kpc and $\approx 10$\% of the total (reddened) flux in the galaxy found beyond the $B$-band isophotal level of $\mu_{\text{B}} = 25$ mag arcsec$^{-2}$. This emission is detected in $7$\% of the individual galaxies, and an additional one third have at least $5$\% of their total flux found beyond $\mu_{\text{B}} = 25$ mag arcsec$^{-2}$ in a disk component. The extraplanar luminosities and colors are consistent with reflection nebulae rather than stellar halos and indicate that, on average, disk galaxies have an extraplanar dust mass of $5\% - 15\%$ of that in their interstellar medium. This suggests that recycled material composes at least a third of the inner circumgalactic medium ($R < 10$ kpc) in $\sim L^{*}$ galaxies.

\end{abstract}



\section{Introduction}\label{sec:intro}

Cosmic dust grains form in star-forming regions and can be dispersed within and around galaxies by star-formation feedback, AGN outflows, and galaxy interactions. The presence and properties of circumgalactic dust (CGD) thus encode a rich history about the circulation of dust-bearing gas between star-forming regions and the circumgalactic medium (CGM). How dust is transported into the CGM by galactic fountains, outflows, and/or tidal stripping, as well as how it survives in the hot halo environment, remain open questions. The dust itself may be a key factor in driving galactic outflows via radiation pressure from reprocessed infrared photons produced when dust grains absorb ultraviolet (UV) light from young stars \citep[e.g.,][]{2011ApJ...735...66M}. Dust grains also play an important role in regulating the physical conditions in dust-bearing gas, affecting the heating and cooling balance, enabling molecule formation, and depleting metals from the gas phase.    

Multi-wavelength observations reveal CGD around the Milky Way and nearby galaxies. In the Galaxy, the chemical abundance patterns of halo gas detected in absorption show significant depletion of refractory elements onto dust grains \citep[e.g.,][]{1996ARA&A..34..279S, 2000ApJ...544L.107W}. On larger spatial scales, \citet{2010MNRAS.405.1025M} use the reddening of background quasars to infer the presence of CGD around external, $\sim L^{*}$ galaxies at $z \approx 0.3$ to impact parameters greater than 1 Mpc in the Sloan Digital Sky Survey. If this reddening is due to CGD, then approximately half of the dust produced in galaxy disks has been redistributed across CGM scales (see also \citealt{2015ApJ...813....7P}, though note the potential contribution of extended disks to the signal as discussed by \citealt{2016MNRAS.462..331S}).

In external galaxies, direct detection of dust grains and polycyclic aromatic hydrocarbons (PAHs) via thermal emission in the far- and mid-infrared reveal CGD entrained in galactic winds \citep[e.g.,][]{2009ApJ...698L.125K, 2010A&A...514A..14K, 2010A&A...518L..66R, 2015ApJ...804...46M, 2018MNRAS.477..699M}. Extraplanar dust is also seen in extinction at the disk-halo interfaces of nearby, normal galaxies \citep[e.g.,][]{2000AJ....119..644H}, suggesting the circulation of dust-bearing gas through galactic fountain flows. Absorption-line spectroscopy indicates diverse origins for CGD, with contributions from fountain flows \citep[e.g.,][]{2019ApJ...876..101Q}, outflows \citep[e.g.,][]{2021MNRAS.502.3733W}, and interactions \citep[e.g.,][]{2022ApJ...926L..33B}.

Reflection nebulae, or scattered starlight produced by dust grains, are a key means by which to study CGD. The UV wavelength regime is well-suited to the detection of these nebulae due to the high scattering cross-section and relatively dark sky. UV reflection nebulae have been detected around tens of nearby ($D \lesssim 25$ Mpc), edge-on disk galaxies, where they occur in both starburst outflows and around more moderately star-forming systems \citep{2005ApJ...619L..99H, 2014ApJ...785L..18S, 2014ApJ...789..131H, 2016ApJ...833...58H, 2018ApJ...862...25J}. These reflection nebulae have UV luminosities that are typically a few percent to $20$\% of the apparent (reddened) UV galaxy luminosity. They are detected on scales as large as $10 - 20$ kpc, with characteristic scale heights of a few kpc. The reflection nebulae around normal galaxies have thick disk morphologies, while more extended and filamentary nebulae are associated with galactic outflows \citep[e.g.,][]{2016ApJ...833...58H}. The luminosities and colors of the nebulae disfavor an origin in stellar halos, and a strong correlation between the extraplanar and disk luminosities is consistent with reflection nebulae \citep{2014ApJ...789..131H, 2016ApJ...833...58H}. 

The small sample sizes of UV reflection nebulae have limited exploration of how CGD properties depend on galaxy luminosity, morphology, and star-formation rate (SFR). The current picture suggests that dust grains are likely found in both ionized and neutral gas phases and that both outflows and interactions contribute to the extraplanar dust layer within a single galaxy \citep[e.g.,][]{2010A&A...514A..14K, 2021MNRAS.502..969Y}. A first step in understanding how CGD depends on host galaxy properties is to explore a larger sample to establish statistically significant relationships.

Here we present a search for UV reflection nebulae around 550 nearby ($D < 100$ Mpc), highly inclined disk galaxies using archival \textit{Galaxy Evolution Explorer} (\textit{GALEX}) near-UV (NUV) and far-UV (FUV) imaging. This expands on previous sample sizes by more than an order of magnitude. We undertake two analysis approaches: first, we search for extraplanar emisssion in all of the sample galaxies individually, and then we repeat the analysis in stacks of galaxies constructed based on galaxy morphology and SFR. We then examine the properties of the extraplanar emission to assess the likelihood that it is produced by reflection nebulae. Throughout this work, we model the effect of the \textit{GALEX} point-spread function (PSF; $4'' - 5''$) to account for contamination by the PSF wings. 

This paper is organized as follows. In \S~\ref{sec:data}, we present the sample selection and the archival data set. We then discuss the data analysis in \S~\ref{sec:analysis}, where we detail the construction and modeling of individual and stacked galaxy luminosity profiles. We present the detection rates and properties of extraplanar emission in the full sample and in the stacks in \S~\ref{sec:results}. In \S~\ref{sec:disc}, we interpret the relative incidence of this emission in the individual galaxies and in the stacks, and we discuss implications for the inferred extraplanar dust mass and the recycled fraction of the inner CGM. Throughout this paper, we assume a $\Lambda$ cosmology with $\Omega_{\text{M}} = 0.3$, $\Omega_{\Lambda} = 0.7$, and $H_{0} = 70$ km s$^{-1}$ Mpc$^{-1}$.

\section{Data}\label{sec:data}

This study uses archival imaging of nearby, edge-on disk galaxies observed in the FUV ($\lambda_{\text{eff}} \approx 1528$ \AA) and NUV band ($\lambda_{\text{eff}} \approx 2310$ \AA) with \textit{GALEX}\footnote{\url{https://archive.stsci.edu/missions-and-data/galex}}. We adopt a model of the average PSF\footnote{\url{http://www.galex.caltech.edu/researcher/techdoc-ch5.html\#2}} obtained from in-flight observations by the \textit{GALEX} team\footnote{There is moderate variation in the PSF with time and across the \textit{GALEX} field of view due to the attitude correction and optics \citep{Morrissey2007}. The potential impact of adopting the nominal (narrowest) PSF is twofold. First, the disk scale heights, which are almost always unresolved, may be overestimated; however, this is the case regardless of the chosen PSF, as discussed in \S~\ref{sec:individual_results} and \S~\ref{sec:stack_results}, and does not affect our ability to detect extraplanar emission on kpc scales. Second, extraplanar emission with small scale heights is more difficult to separate from the disk at the signal-to-noise ratio of the individual galaxies. Therefore, adopting the nominal PSF is a conservative approach which may underestimate the detection rate of compact extraplanar emission. We further discuss the impact of PSF variation in \S~\ref{sec:psf}.}. The full width at half maximum of the \textit{GALEX} PSF is $4.3''$ in the FUV and $5.3''$ in the NUV, or $2$ kpc and $2.5$ kpc at $D = 100$ Mpc, the maximum distance in the sample (see \S~\ref{sec:sample}). A radius of $7''$ and $10''$ encloses 90\% of the energy in the FUV and NUV, or $3.2$ kpc and $4.6$ kpc at $D = 100$ Mpc, respectively. The PSF profile has a roughly Gaussian core with extended wings, including a plateau in the NUV PSF at about the $0.1$\% level out to $\approx 45''$. \textit{GALEX} has a field of view of $1.25^{\circ}$ and a plate scale of $1.5''$ pixel$^{-1}$. The \textit{GALEX} Medium Imaging Survey (MIS) observed $1000$ square degrees of sky with sufficient depth ($t_{\text{exp}} = 1500$ s) to probe faint extraplanar emission, and we draw primarily from the MIS to construct the sample here.

\subsection{Sample Selection}\label{sec:sample}

To detect extraplanar emission, we require a signal-to-noise ratio (S/N) of $\text{S/N} \gtrsim 3$ per pixel in the one-dimensional vertical luminosity profiles constructed in \S~\ref{sec:analysis}, to allow robust model fitting to the planar and extraplanar emission. We estimate the expected S/N by adopting the median absolute magnitudes ($M_{\text{NUV}} = -13.7$ mag and $M_{\text{FUV}} = -12.9$ mag) of the extraplanar emission in 26 nearby, highly inclined galaxies ($M_{\text{B}} < -20$ mag) observed by \citet{2016ApJ...833...58H}. We also assume that the emission is uniformly distributed over a region with a length of $10$ kpc and a width equal to twice the median scale height measured by \citet{2016ApJ...833...58H}, $h_{\text{h}} = 3.4$ kpc. This yields an estimated surface brightness of $\mu_{\text{NUV}} = 27.5$ mag arcsec$^{-2}$ and $\mu_{\text{FUV}} = 28.3$ mag arcsec$^{-2}$. This is comparable to the typical \textit{GALEX} background ($\mu_{\text{NUV}} = 27.4$ mag arcsec$^{-2}$ and $\mu_{\text{FUV}} = 28.0$ mag arcsec$^{-2}$).

We estimate the S/N per pixel in the two-dimensional image, accounting for Poisson noise from the source and background. We then determine the S/N per pixel in the one-dimensional profile constructed by averaging over the dimension parallel to the major axis (see \S~\ref{sec:prof_proc}), assuming an improvement of $\sqrt{N_{\text{pix}}}$, where $N_{\text{pix}}$ is the number of pixels. This yields a detection significance of $\approx 4\sigma$ in the NUV at a distance of $D = 100$ Mpc with an exposure time of $t_{\text{exp}} = 1000$ s. A similar detection significance ($\approx 3\sigma$) is found at a distance of $D = 50$ Mpc in the FUV with the same exposure time. There is evidence that the extraplanar luminosity is positively correlated with the galaxy luminosity \citep{2016ApJ...833...58H}, which implies $\approx 2\sigma$ halo detections in the NUV down to $M_{\text{B}} \approx -19$ mag at comparable distances and exposure times.

From the HyperLeda database\footnote{\url{http://leda.univ-lyon1.fr/}} \citep{2014A&A...570A..13M}, we select late-type galaxies (Sa - Sd) with $M_{\text{B}} < -19$ mag and $D \le 100$ Mpc; we also require an inclination angle $i \gtrsim 75^{\circ}$ to allow for spatial separation of the planar and extraplanar regions. From this initial sample, we identify galaxies with \textit{GALEX} imaging with $t_{\text{exp}} \ge 1000$ s in either the FUV and/or NUV bands. The median exposure time of the \textit{GALEX} All-Sky Imaging Survey is below our exposure time cutoff, and thus the majority of the data used in this study are from the Medium Imaging Survey, the Nearby Galaxy Survey, and guest observer programs. We exclude by eye galaxies that are morphologically disturbed, including warps. We also exclude those that have nearby, bright stars that require masking a significant portion of the galaxy, or have positions and/or angular sizes that do not permit the extraction box (see below) to fit within the \textit{GALEX} field of view. The final sample consists of 551 galaxies, of which 343 (546) have data available in the FUV (NUV) band. There are 338 galaxies that have data available in both bands. All of the \textit{GALEX} data used in this study can be found in MAST: \dataset[https://doi.org/10.17909/c1xr-dy81]{https://doi.org/10.17909/c1xr-dy81}. We present the properties of the galaxy sample in Appendix~\ref{sec:gal_sample}.

\subsection{Image Preparation}\label{sec:image_prep}

For each galaxy, we obtain a pipeline-reduced intensity map from the \textit{GALEX} data release GR6/7 via the public archive\footnote{\url{http://galex.stsci.edu/GR6/\#releases}}. For galaxies with intensity maps available from more than one observation, we select the map with the longest exposure time (stacking between multiple observations of the same galaxy is not done to avoid source blurring due to astrometric effects and PSF distortions). We crop each map to a $0.11$ square degree image centered on the galaxy. We then use SExtractor\footnote{\url{https://www.astromatic.net/software/sextractor/}} \citep{1996A&AS..117..393B} to identify and mask point sources in the field, where each object is masked with a circular aperture with radius equal to the Kron radius. The point source masks were inspected by eye, and we perform additional masking by hand where necessary, including in the case of extended sources in the foreground or background.

We do not mask any sources within a box around the galaxy to prevent masking star-forming regions and other intrinsic structures within the galaxy. We define the box to have length equal to the major axis at the $B$-band isophotal level of $\mu_{\text{B}} = 25$ mag arcsec$^{-2}$, $D_{\text{25}}$, and width equal to the minor axis, $B_{\text{25}} = D_{\text{25}}R_{\text{25}}$. We obtain $D_{\text{25}}$ and $R_{\text{25}}$ from HyperLeda. We denote half of the major and minor axes as $r_{\text{25}} = D_{\text{25}}/2$ and $b_{\text{25}} = B_{\text{25}}/2$, respectively.

\section{Analysis}\label{sec:analysis}

We conduct a two-stage analysis to characterize the presence and properties of extraplanar emission in the UV. In the first stage, we analyze the vertical luminosity profiles of all individual galaxies. In the second stage, we stack the luminosity profiles by morphological type and SFR to improve the detection significance of the faint extraplanar light. We also explore relationships between the scale height and luminosity of the extraplanar emission and the properties of the underlying galaxies. Here, we discuss the creation of the luminosity profiles for the individual (\S~\ref{sec:prof_proc}) and stacked galaxies (\S~\ref{sec:stack_proc}) and the model fitting procedure used to assess the presence of extraplanar light (\S~\ref{sec:fit_proc}).

\subsection{Vertical UV Luminosity Profiles}\label{sec:prof_proc}

For each galaxy, we construct the vertical (perpendicular to the disk) UV luminosity profiles for both \textit{GALEX} bands. The motivation for analyzing one-dimensional profiles is to improve the S/N of the faint extraplanar emission and to average over clumpiness in the planar and (potentially) extraplanar components. We define an extraction box over which to construct the profile that has a width equal to $D_{\text{25}}$ and a height that is twice the width. This choice of box height captures at least several times the expected scale height of the extraplanar emission \citep[e.g.,][]{2016ApJ...833...58H}. Within the box, we calculate the median intensity as a function of height above or below the disk, $z$, excluding the masked pixels. Finally, we calculate the corresponding uncertainty due to Poisson noise and background fluctuations, where the latter is the standard deviation of the median intensity at $|z| > b_{\text{25}}$.

\subsection{Stacking Procedure}\label{sec:stack_proc}

To investigate the dependence of extraplanar emission on galaxy properties, we stack the vertical luminosity profiles based on galaxy morphology and SFR. For the morphology criterion, we define four bins from early to late morphological type: Sa, Sb, Sc, and Sd galaxies. We include both barred and unbarred galaxies in the stacks, and we exclude galaxies with intermediate classifications (e.g., Sab; this is approximately one third of the sample). For the SFR criterion, we define five bins: $\text{SFR} \le 0.1$, $0.1 < \text{SFR} \le 0.3$, $0.3 < \text{SFR} \le 1$, $1 < \text{SFR} \le 3$, and $\text{SFR} > 3$, where the SFR has units of $M_{\odot}~\text{yr}^{-1}$. These bins are chosen to represent the dynamic range of the sample while providing sufficient statistics in each stack. The number of galaxies included in each stack are given in Tab.~\ref{tab:stack_results}.

To construct each stack, we first project the profiles to a common distance of $D_{\text{max}} = 100$ Mpc, the limiting distance of the sample. A careful treatment of the PSF is needed to stack the projected galaxies. Since the galaxies are observed at a range of distances, they have a range of effective PSFs when projected to a common distance. For example, for galaxies observed at $D = 25$ Mpc and $D = 100$ Mpc, the NUV PSF of $5.3''$ corresponds to a physical scale of $0.6$ kpc and $2.5$ kpc, respectively. Consequently, when the galaxy observed at $D = 25$ Mpc is projected to $D = 100$ Mpc, it has an effective PSF that is a factor of a few better than if the galaxy had been observed at $D = 100$ Mpc originally. A sample of galaxies projected to $D_{\text{max}}$ thus has a range of effective PSFs, and a stack of these galaxies cannot be well fit with a single, common PSF. To address this, we convolve the profiles with the \textit{GALEX} PSF, including both the PSF core and wings, after projecting them to $D_{\text{max}}$, to approximate observing them at this distance originally.

As the profiles are now twice convolved with the PSF, they are modestly artificially broadened; closer galaxies, which are less impacted by PSF convolution at the time of observation, experience less artificial broadening than those farther away. Due to the double convolution, the measured scale heights of all emission components approximates the true scale height convolved with the PSF at the original (observed) distance. This sets a floor in the minimum measurable scale height that is determined by the physical scale (in kpc) of the PSF averaged over the distance distribution. At the median distance of the sample ($D_{\text{med}} = 71$ Mpc), the standard deviation of the PSF (FWHM/($2\sqrt{2\ln{2}}$)) is $\sigma = 0.6$ kpc and $\sigma = 0.7$ kpc in the FUV and NUV, respectively. This exceeds the expected scale height of the stellar thin disk ($h_{\text{d}} \approx 0.2 - 0.4$ kpc; e.g., \citealt{2013ApJ...773...45S, 2016ARA&A..54..529B}), but is significantly smaller than the typical scale height of extraplanar emission produced by reflection nebulae ($h_{\text{h}} \approx 3.5$ kpc; \citealt{2016ApJ...833...58H}). Consequently, we cannot recover the true scale heights of the stellar disks in the galaxy stacks (see discussion in \S~\ref{sec:stack_results}), but we anticipate robust recovery of the scale height and luminosity of any extraplanar emission. We note that the PSF wings are broad compared to the halo scale heights, but the luminosity in the wings typically falls significantly below the detected halo luminosity (see, e.g., Fig.~\ref{fig:stack_Hubble_Sd}), and therefore the wings do not significantly impact the measured halo scale heights.

After convolving the projected profiles with the \textit{GALEX} PSF, we construct the stacks by subtracting a linear model of the background and spatially aligning the profiles by their flux peaks (the background and peak locations are determined by the fitting performed in \S~\ref{sec:fit_proc}). We then calculate a weighted average of the projected, convolved profiles, where the weights are determined by the exposure times ($w_{i} = t_{\text{exp},i}/\sum\limits_{i} t_{\text{exp},i}$, where $t_{\text{exp},i}$ is the exposure time of the $i$th galaxy). We calculate $b_{\text{25}}$ for the stacks by determining the weighted average of the $b_{\text{25}}$ values of the contributing galaxies, using the same weights.

\subsection{Fitting Procedure}\label{sec:fit_proc}

\begin{figure*}
\centering
\includegraphics[scale = 0.6]{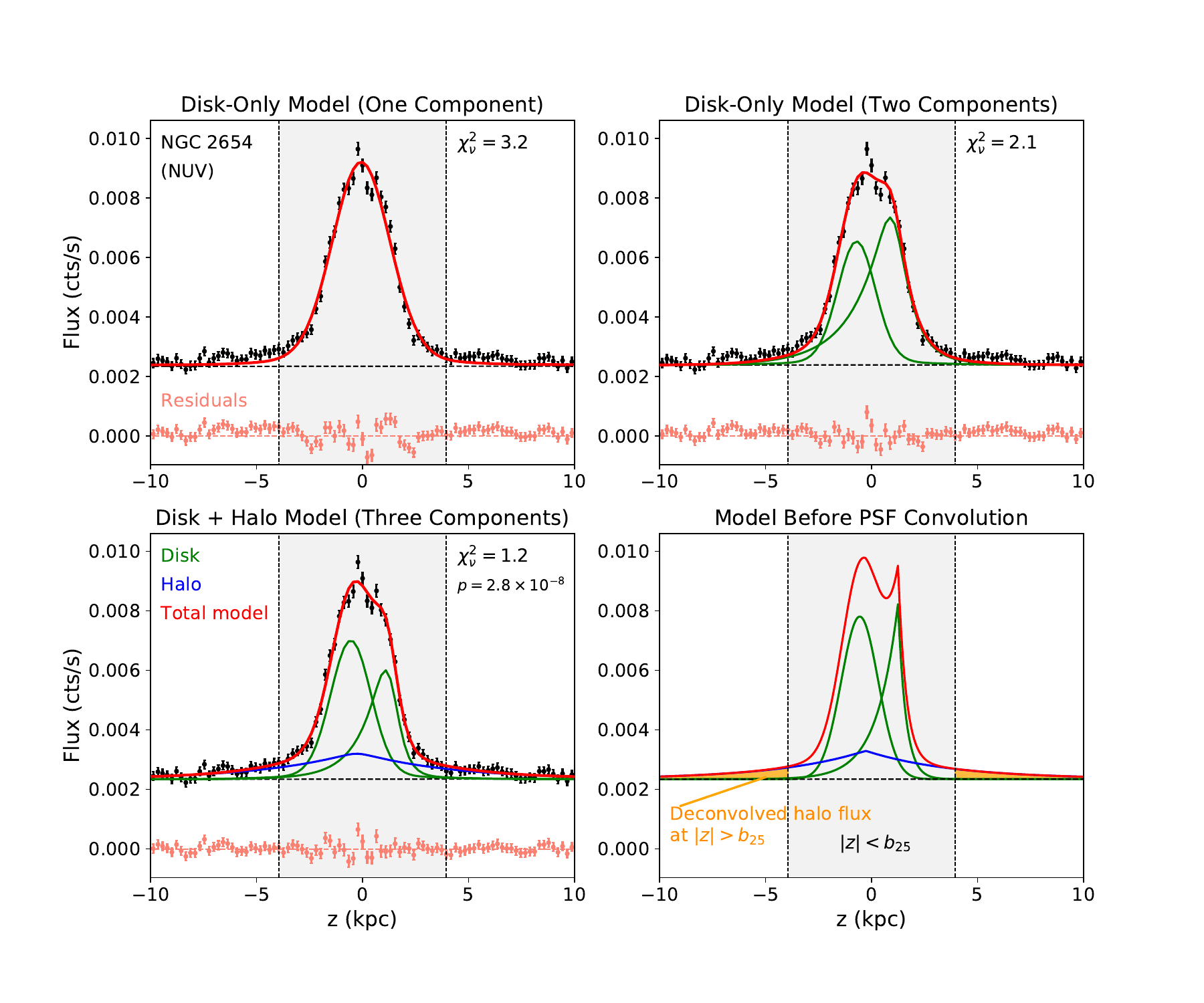}
\caption{Summary of the procedure used to characterize UV halos, shown here for the example galaxy NGC 2654 in the NUV. We first fit a disk-only model with a single component to the vertical luminosity profile (upper left). We then add a second disk component if this sufficiently improves the quality of the fit ($p \leq 0.05$; upper right). Next, we add the halo component, again if sufficient improvement is seen ($p \leq 0.05$; lower left). We consider the extraplanar luminosity to be the emission in the deconvolved halo at $|z| > b_{25}$ ($\mu_{\text{B}} = 25$ mag arcsec$^{-2}$), as shown at lower right; all emission within $|z| \leq b_{25}$ is considered stellar light.}
\label{fig:ngc2654}
\end{figure*}

We fit each vertical luminosity profile to determine whether extraplanar emission is present and measure its scale height and luminosity. We begin by fitting the profile with only a disk component, and then we determine whether including extraplanar emission, which we refer to as the halo component, significantly improves the fit. The summarized steps are:
\begin{enumerate}
\item Fit a disk-only model with one component (Gaussian or exponential; \S~\ref{sec:fit_proc_indiv}).
\item Add a second component to the disk model (Gaussian or exponential); determine whether the second component significantly improves the fit using an F-test ($p$-value $\leq 0.05$; \S~\ref{sec:fit_proc_indiv}).
\item Add an exponential halo component to the chosen disk model; determine whether the halo significantly improves the fit, again using an F-test ($p \leq 0.05$; \S~\ref{sec:fit_proc_indiv_halo}).
\item If the halo model is selected and both FUV and NUV data are available, fit the halo simultaneously in both bands for improved constraints on the scale height and amplitude (\S~\ref{sec:fit_proc_indiv_halo}).
\item For all detected halos, measure the deconvolved halo emission at $|z| > b_{25}$; we adopt this as the extraplanar luminosity (\S~\ref{sec:quant_halo}).
\end{enumerate}
These steps are illustrated for an example galaxy in Fig.~\ref{fig:ngc2654} and are discussed in more detail in the sections below. For all model fits, we include a linear fit to the background.

There are two complexities to address in the fitting. The first is to properly account for the intrinsically two-dimensional PSF when fitting one-dimensional profiles. The amount of emission that is scattered into the extraplanar region by the convolution depends on the scale length of the galaxy, which is not preserved when flattening the two-dimensional images into one dimension. Therefore, the models must be convolved with the PSF in two dimensions. We first construct a two-dimensional model, convolve this model with the PSF, and create a one-dimensional profile following the same masking and extraction procedures used for the data (\S~\ref{sec:prof_proc}).

Secondly, the luminosity profiles of the disks are intrinsically complex. The UV traces young, clumpy stellar populations and is highly susceptible to extinction, resulting in more irregular profiles than in the optical or near infrared. In principle, we expect the disks to be well fit by a single exponential distribution. However, in practice, the UV disks can depart from single exponentials due to structure in the stellar population, including multiple (e.g., thin and thick) disks, clumping of star-forming regions, and disk perturbations such as warps. Additionally, the disk emission is modified by central dust lanes and other anisotropic extinction. The PSF often prevents disentangling which of these effects are at play. This motivates testing disk models of different forms (exponential and Gaussian) and number of components (one or two disk components), as discsused below. There may also be intrinsic differences between the disks in the FUV and the NUV (the former of which traces younger stars), depending on the ages and spatial distributions of the stellar populations. We thus fit the FUV and NUV profiles separately. We use the IDL routine \texttt{mpfitfun}\footnote{\url{https://www.l3harrisgeospatial.com/docs/mpfitfun.html}} to perform the fit, which employs a Levenberg-Marquardt least-squares minimizer. We report the best-fit values and the errors determined from the diagonal terms of the covariance matrix returned by \texttt{mpfitfun}. In the following sections, we provide details on the model fitting for individual (\S~\ref{sec:fit_proc_indiv} and \S~\ref{sec:fit_proc_indiv_halo}) and stacked galaxies (\S~\ref{sec:fit_proc_stacks}).

\subsubsection{Fitting Individual Galaxies - the Disk}\label{sec:fit_proc_indiv}

The complexities discussed above require us to diversify the disk model set beyond single exponential disks. We select this model set by empirically determining the simplest distributions that well fit the disk profiles. We first consider single exponential and Gaussian profiles; the former nominally describes a smooth disk, while the latter describes a clumpier stellar population. We find that approximately half of the sample galaxies are better fit by a Gaussian profile than by an exponential. We thus allow for use of a Gaussian profile if it provides an improved fit compared to an exponential.

We find that the disks are not well fit by a single component in approximately one third of the galaxies. We therefore allow for a second disk component if it significantly improves the fit using an F-test as described below. We add either an exponential or a Gaussian component, depending on which better improves the fit. We thus arrive at disks modeled by two exponentials (i.e., two smooth disks), an exponential and a Gaussian (a smooth disk modified by clumping), or two Gaussians (a highly clumpy distribution).

We now describe the fitting procedure quantitatively. The exponential disk is of the form:
\begin{equation}
    f(z)_{\text{d}} = \begin{cases}
    f_{0,\text{d}}e^{-\frac{|z|}{h_{\text{d1}}}} &\text{if $z \le z_{0}$}\\
    f_{0,\text{d}}e^{-\frac{|z|}{h_{\text{d2}}}} &\text{if $z > z_{0}$},
    \end{cases}
\end{equation}
where $z$ is the distance from the midplane, $z_{0}$, and $f_{0,\text{d}}$ is the flux at $z_{0}$. The exponential scale heights, $h_{\text{d1}}$ and $h_{\text{d2}}$, are allowed to differ on each side of the disk. This accounts for projection effects in galaxies at lower inclination angles. The Gaussian takes the form:
\begin{equation}
    f(z)_{\text{d}} = f_{0,\text{d}}e^{-z^{2}/(2h_{\text{d}}^{2})},
\label{eq:gauss_disk}
\end{equation}
where $h_{\text{d}}$ is the scale height. As the Gaussian distribution is intended to describe clumping rather than a smooth disk, we do not allow the scale height to vary on each side of the disk.

As shown in the top left panel of Fig.~\ref{fig:ngc2654}, we first select either a single exponential or a single Gaussian model based on the fit with the lowest $\chi_{\nu}^{2} = \chi^{2}/\nu = \sum\limits_{i} ((f_{i,\text{obs}} - f_{i,\text{mod}})^2/\sigma_{i}^{2})/\nu$. Here, $f_{i,\text{obs}}$ and $f_{i,\text{mod}}$ are the observed and modeled fluxes in the $i$th spatial bin, respectively, and $\sigma_{i}$ is the uncertainty on the observed quantity. The number of degrees of freedom, $\nu = N - m$, is the difference between the number of spatial bins, $N$, and the number of free parameters, $m$. All quality-of-fit statistics are computed over $|z| \leq 2b_{\text{25}}$.

Next, as shown in the top right panel of Fig.~\ref{fig:ngc2654}, we assess whether an additional disk component is required using an F-test. We fit two, two-component disk models by adding an exponential or a Gaussian component to the original best-fit model; the parameters of both the first and second components are allowed to vary in the two-component fit. We then calculate an F-statistic of the form:
\begin{equation}
F = ((\chi_{1}^{2} - \chi_{2}^{2})/(m_{2} - m_{1}))/\chi_{\nu,2}^{2}.
\label{eq:fstat}
\end{equation}
Here, the subscripts refer to the $\chi^{2}$, $\chi_{\nu}^{2}$, and $m$ values for the one- (1) and two-component models (2). We then compute the F-test probability, $p$; this quantifies the likelihood that improvement in the model fit due to the addition of the second component occurs by chance. If $p \le 0.05$ for either of the two-component models, we adopt this model as the best fit. In the case that this condition is met for both models, we select the model with the lower $p$ value. 

To ensure that the second disk component is physically motivated, we impose several priors on the fit. The second component is intended to characterize structure in the disk rather than emission on halo scales. Thus, we set the prior that the scale heights of both disk components do not exceed the scale height determined from the one-component model (the prior is imposed based on the maximum scale height in the case of the asymmetric exponential profile). We also intend for the weaker component to be sufficiently strong to reflect true sub-structure, and we thus set a prior that the amplitude of the second component is at least $10\%$ of the amplitude of the one-component model. The centroids of both components are allowed to vary from $z_{0}$ above. To robustly detect extraplanar emission, it is more important to appropriately assign the disk emission to the disk component(s) than to recover the intrinsic properties (e.g., scale height) of the disk. We emphasize that the goal is to select the simplest model that well characterizes the disk distribution, so that the presence of extraplanar light can be assessed.

\subsubsection{Fitting Individual Galaxies - the Halo}\label{sec:fit_proc_indiv_halo}

We next test whether a halo component is present (see the bottom left panel of Fig.~\ref{fig:ngc2654}). We consider only an exponential profile for the halo. This is empirically motivated by our sample and by past studies of UV halos, which find that the halo morphologies resemble thick disks \citep[e.g.,][]{2016ApJ...833...58H}. The profile has the form:
\begin{equation}
f(z)_{\text{h}} = f_{0,\text{h}}e^{-\frac{|z - z_{0,\text{h}}|}{h_{\text{h}}}}.
\label{eq:gauss_halo}
\end{equation}
Here, $f_{0,\text{h}}$ is the halo flux at $z_{0,\text{h}}$ and $h_{\text{h}}$ is the halo scale height. We impose several priors to ensure that the halo is a sub-dominant component that is more spatially extended than the disk. We do not allow $f_{0,\text{h}}$ to exceed half of the amplitude of the best-fit one-component disk model or $h_{\text{h}}$ to be less than the maximum scale height selected by the best-fit (one- or two-component) disk model. We assume that the midpoint of the halo, $z_{0,\text{h}}$, lies in the midplane of the galaxy. For a one-component disk model, we assume that the disk midpoint reflects the true midplane, and we tie the midpoints of the disk and the halo together ($z_{0,\text{h}}$ = $z_{0}$). For a two-component disk model, neither of the disk midpoints may reflect the true midplane if they capture clumpy sub-structure. We thus require the halo midpoint to fall within 10\% of the observed FWHM of the profile peak.

We expect UV halos to have the same scale height in the FUV and NUV. For galaxies with data available in both bands, we can more precisely measure the halo scale heights and luminosities by imposing this restriction than from either band alone. For these galaxies, if a halo is detected in at least one band, we re-perform the fit in the FUV and NUV simultaneously on a grid in halo scale height ($h_{\text{h}} = 0.5 - 8.0$ kpc, in steps of $0.5$ kpc) and fractional NUV flux in the halo ($f_{\text{h,NUV}} = 0.05 - 0.95$, in steps of 0.05). From dust models \citep[e.g.,][]{2001ApJ...548..296W}, we expect reflection nebulae to be either bluer than the disks or comparable in color, and we require the fractional flux in the FUV to match or exceed that in the NUV. We adopt the the best-fit disk model as determined for each band in \S~\ref{sec:fit_proc_indiv}, but we allow the disk parameters to vary between grid points. 

Assuming a likelihood of the form $\mathcal{L} \propto e^{-\chi^{2}/2}$, we construct marginalized probability distribution functions for the halo scale heights and fractional fluxes by calculating a joint $\chi^{2}$ over both bands. From the cumulative distribution functions, we report the median as the best-fit value and the $68$\% confidence interval as the $1\sigma$ uncertainties. For both the single-band and two-band fits, we again assess the need for the halo component by calculating the F-statistic (Eq.~\ref{eq:fstat}) and corresponding $p$-value. In addition to requiring $p \le 0.05$ to adopt the halo model, we also require that the best-fit value of $f_{\text{h,NUV}}$ exceed zero at the $\ge 2\sigma$ level. The latter removes cases of weak detections; visual inspection indicates that at least some of these cases are due to non-linear structure in the background. These conservative criteria favor fidelity of the halo detections at the potential expense of the loss of a small number of weak detections.

\subsubsection{Fitting Stacked Galaxies}\label{sec:fit_proc_stacks}

We follow the procedures in \S~\ref{sec:fit_proc_indiv} and \S~\ref{sec:fit_proc_indiv_halo} to fit a simplified set of models to the vertical luminosity profiles of the galaxy stacks. The disks of the stacked profiles are well described by a single Gaussian, with a $\chi^{2}$ test preferring a Gaussian disk to an exponential disk for all stacks. This is a result of the second PSF convolution performed after projecting the profiles to $D_{\text{max}}$ (see \S~\ref{sec:stack_proc}); due to the double convolution, the vertical profile of the disk approximates the true profiles convolved with the Gaussian core of the PSF at the original (observed) distances. We thus fit a single Gaussian disk model (Eq.~\ref{eq:gauss_disk}), refit adding an exponential halo (Eq.~\ref{eq:gauss_halo}), and assess the need for the halo component using the F-test (Eq.~\ref{eq:fstat}). We apply the same halo priors and detection criteria as for the individual galaxies.

We use bootstrap resampling to determine the probability distributions of the model parameters. For each bin containing $N_{\text{gal}}$ galaxies, we resample the galaxies $10^{3}$ times, drawing $N_{\text{gal}}$ galaxies each time with replacement. We then construct the galaxy stack and perform the fit as described above for each resampled set of galaxies. For each model parameter, we construct the probability distribution from the best-fit values determined from each fit, and we report the median value and the $68$\% confidence interval as the corresponding $1\sigma$ uncertainties. The halo detection rates and scale heights are consistent between the bands within the uncertainties, and thus we do not perform the multi-band grid search for the galaxy stacks.

\subsubsection{Quantifying Extraplanar Emission}\label{sec:quant_halo}

The goal of this analysis is to quantify the UV luminosity of extraplanar emission. This requires defining ``extraplanar''. As our objective is to identify scattered light from a dusty disk-halo interface and/or CGM, we seek to separate scattered light as cleanly as possible from starlight. The model profile fitting provides a starting point by separating the light into disk and halo components. However, a significant fraction of the light in the halo component is co-spatial with the disk, and this light may come from the stellar population rather than from a reflection nebula. We therefore define a distance from the disk, $|z| = b_{25}$, which separates planar and extraplanar emission, where $b_{25}$ is defined as in \S~\ref{sec:image_prep}. We consider all light in the halo component found at $|z| > b_{\text{25}}$ to be extraplanar emission and all other light to be starlight. We illustrate this distinction in the bottom right panel of Fig.~\ref{fig:ngc2654}. Throughout this work, the term ``halo'' is used to refer to the model component defined in \S~\ref{sec:fit_proc_indiv_halo}, while ``extraplanar emission'' is used to refer to the light in this component found at $|z| > b_{\text{25}}$.

$b_{25}$ is chosen to conservatively enclose the thin and thick stellar disks within $|z| < b_{25}$. The median value of $b_{25}$ in the sample, $b_{25} \approx 3$ kpc, is a factor of $\gtrsim 6$ and $\sim 3$ times the typical thin- and thick-disk scale heights in an $L^{*}$ galaxy, respectively. In \S~\ref{sec:individual_results}, we show that only a few percent of the light in the disk component is typically found at $|z| > b_{\text{25}}$, demonstrating that this is a reasonable demarcation between starlight and extraplanar emission. It is possible that significant scattered light is found within $|z| < b_{25}$, and thus this criterion provides a conservative measure of the extraplanar emission in order to ensure minimal contamination by starlight.

\section{Results}\label{sec:results}

\subsection{Individual Galaxy Results}\label{sec:individual_results}

\begin{figure*}
\centering
\includegraphics[scale = 0.75]{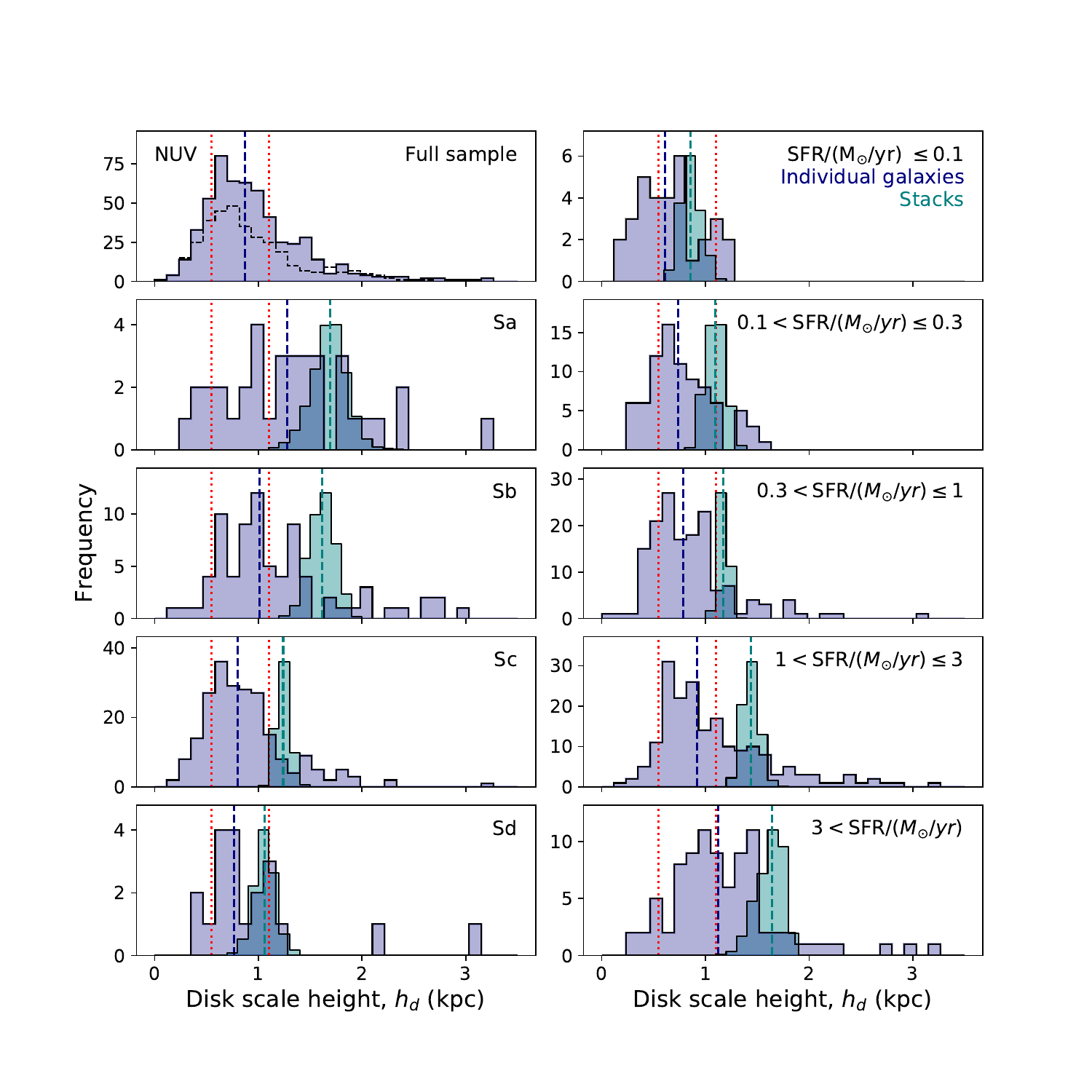}
\caption{Measured disk scale heights are broadened beyond typical thin disk scale heights due to clumpy UV morphologies, inclined disks ($i < 90^{\circ}$), and the \textit{GALEX} PSF. The NUV disk scale heights for the individual galaxies and the stacks are shown in blue and cyan, respectively; the median is given by the dashed lines. The full sample is displayed at the upper left; the FUV distribution is shown as the dashed histogram for comparison. We separate the sample by morphological type in the remainder of the left column and by SFR in the right column. The stack distributions result from bootstrap resampling and are renormalized for display purposes. The dotted red lines indicate the NUV PSF (FWHM/$2\sqrt{2\ln{2}}$) at $D = 50$ Mpc and $D = 100$ Mpc. See Appendix \ref{sec:supp_fuv} for the equivalent figure for the FUV data.}
\label{fig:hgal_nuv}
\end{figure*}

We first consider evidence for extraplanar emission from the individual galaxies. We report the disk properties in Tab.~\ref{tab:disks}, including the best-fit disk model, flux, fractional flux at $|z| > b_{25}$, scale height(s), and FUV - NUV color. The observed and modeled profiles of three example galaxies are shown in Appendix~\ref{sec:lum_prof}. The best-fit model for $37$\% of the sample in the NUV ($31$\% in the FUV) requires two disk components, demonstrating that clumpiness in the stellar population is relatively common in the sample.

We show the distributions of the disk scale heights for the full sample in the top panel of Fig.~\ref{fig:hgal_nuv} (for galaxies with asymmetric profiles and/or multiple components, the median scale height is shown). The scale heights are generally consistent between the FUV and NUV within the errors. The median scale height in the NUV (FUV) is $h_{\text{d}} = 0.9$ kpc ($h_{\text{d}} = 0.8$ kpc), with $68$\% of the sample ranging from $h_{\text{d}} = 0.5 - 1.4$ kpc ($h_{\text{d}} = 0.5 - 1.3$ kpc). The somewhat smaller scale heights in the FUV may reflect the fact that the youngest stars are generally found in the thinnest disks. However, the median value is significantly larger than the stellar thin disk scale heights typically observed in the Milky Way and nearby, resolved galaxies ($h_{\text{d}} \approx 0.2 - 0.4$ kpc).

Several effects contribute to measuring unphysically large disk scale heights. First, the scale height measurement assumes a perfectly edge-on disk, whereas our sample has a range of inclination angles ($i \gtrsim 75^{\circ}$). The measurement may also be inflated by clumpiness in the stellar population and other morphological perturbations, such as those induced by interactions. Additionally, the projected size of the PSF (FWHM/$2\sqrt{2\ln{2}}$ $\approx 0.5$ kpc at $D = 50$ Mpc and $\approx 1.0$ kpc at $D = 100$ Mpc in the NUV) makes it difficult to measure the scale height of an unresolved disk even when accounting for PSF convolution in the model profiles. 
 
In Fig.~\ref{fig:hgal_nuv}, we separate the disk scale heights into bins by morphology and SFR for the NUV (see Appendix \ref{sec:supp_fuv} for the equivalent figure for the FUV). A slight evolution is seen in the scale height with morphological class, with the thinnest disks observed in the latest type galaxies, as expected. Additionally, there is a weak trend toward larger scale heights with increasing SFR; this trend is also expected, as higher SFRs are typically found in more massive disks. However, the PSF causes the measured disk scale heights to be biased high for all morphological types and SFRs, due to the difficulty of resolving disks with scale heights significantly smaller than the PSF.

\begin{figure*}
\centering
\includegraphics[scale = 0.75]{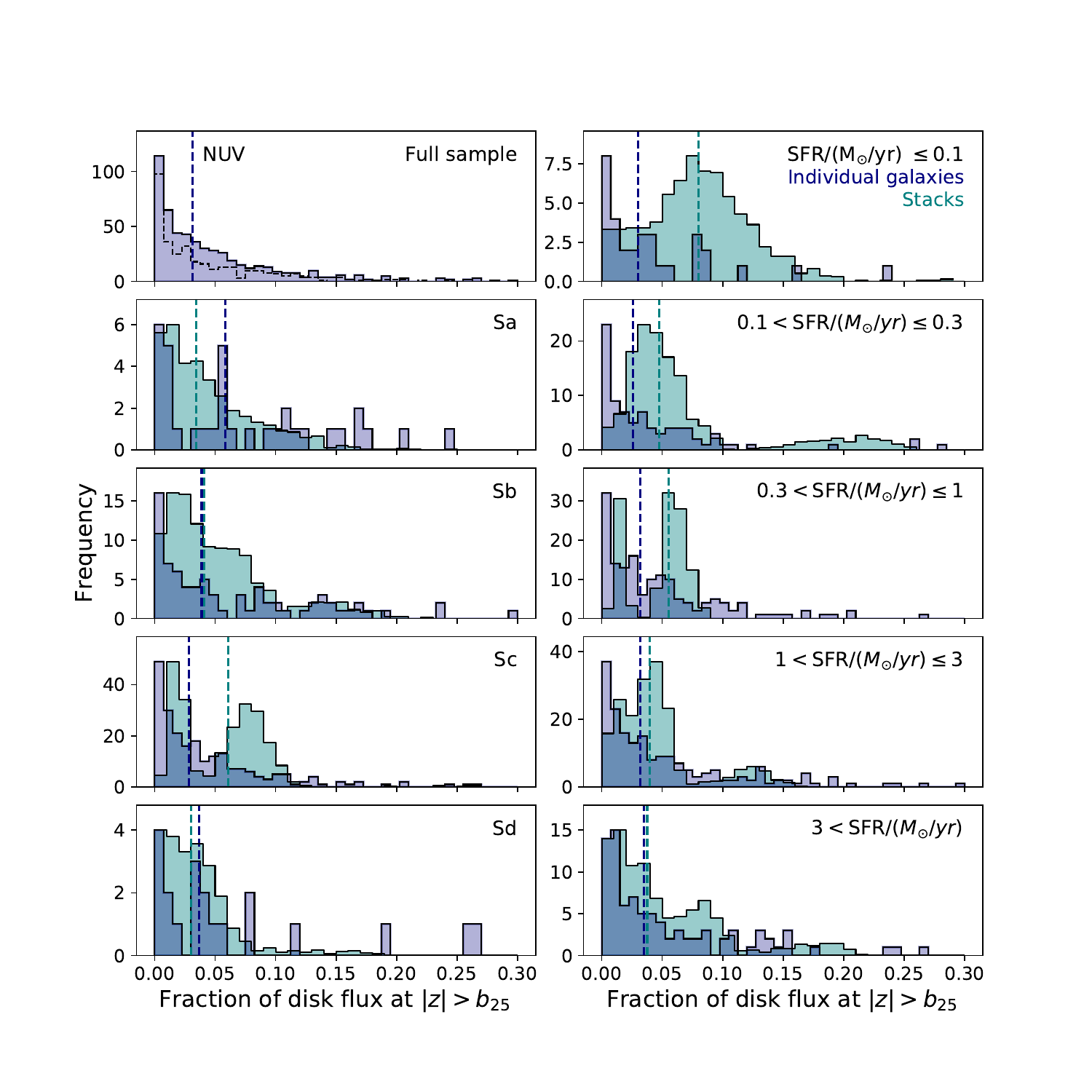}
\caption{Fractional disk flux at $|z| > b_{\text{25}}$, $f_{\text{d,out}}'$, is typically a few percent for the individual galaxies (blue) and the stacks (cyan), demonstrating that $b_{\text{25}}$ reasonably separates planar and extraplanar emission. The stacked distributions are bootstrap resampled. The colors and linestyles are the same as in Fig~\ref{fig:hgal_nuv}. See Appendix \ref{sec:supp_fuv} for the equivalent figure for the FUV data.}
\label{fig:f_frac_nuv}
\end{figure*}

\begin{deluxetable*}{ccccccccc}
  \tablecaption{Disk properties of the sample galaxies}
  \tablehead{
    \colhead{(1)} &
    \colhead{(2)} &
    \colhead{(3)} &
    \colhead{(4)} & 
    \colhead{(5)} &
    \colhead{(6)} &
    \colhead{(7)} &
    \colhead{(8)} &
    \colhead{(9)} \\
    \colhead{Galaxy} &
    \colhead{Disk model} &
    \colhead{$F_{\text{d}}$} &
    \colhead{$f_{\text{d},\text{out}}'$} &
    \colhead{$h_{z,1}$} & 
    \colhead{$h_{z,2}$} &
    \colhead{$h_{z,3}$} &
    \colhead{$h_{z,4}$} &
    \colhead{FUV - NUV} \\
    \colhead{} &
    \colhead{} &
    \colhead{(mJy)} &
    \colhead{} &
    \colhead{(kpc)} & 
    \colhead{(kpc)} &
    \colhead{(kpc)} &
    \colhead{(kpc)} &
    \colhead{(mag)}
  }
  \startdata
  \noalign{\smallskip}
  \multicolumn{9}{c}{\textbf{NUV}} \\
  \smallskip
   PGC000012 & 1G & $0.64 \pm 0.02$ & $0.08$ & $2.37 \pm 0.06$ & - & - & - & $0.45 \pm 0.04$ \\
   PGC000192 & 1G & $0.520 \pm 0.008$ & $0.009$ & $1.49 \pm 0.03$ & - & - & - & - \\
   UGC00043 & 1G & $0.203 \pm 0.007$ & $0.05$ & $1.13 \pm 0.06$ & - & - & - & $0.44 \pm 0.06$ \\
   ... & ... & ... & ... & ... & ... & ... & ... & ... \\
  \noalign{\smallskip}
  \multicolumn{9}{c}{\textbf{FUV}} \\
  \smallskip
  PGC000012 & GE & $0.40 \pm 0.01$ & $0.03$ & $1.70^{+0.6}_{-1.0}$ & - & $0.84^{+1.5}_{-0.7}$ & $1.3 \pm 0.6$ & $0.45 \pm 0.04$ \\
  PGC000192 & -  & - & - & - & - & - & - & - \\
  UGC00043 & 1E & $0.129 \pm 0.006$ & $0.22$ & $1.5 \pm 0.2$ & $0.8 \pm 0.1$ & - & - & $0.44 \pm 0.06$ \\   
   ... & ... & ... & ... & ... & ... & ... & ... & ... \\
  \label{tab:disks}
  \enddata
  \tablecomments{(1) Galaxy name, (2) disk model; 1G = one Gaussian component, 1E = one exponential component, 2G = two Gaussian components, 2E = two exponential components, GE = one Gaussian and one exponential component, (3) total disk flux, (4) fraction of the disk flux at $|z| > b_{\text{25}}$, $f_{\text{d,out}}' = F_{\text{d,out}}/F_{\text{d}}$, (5) - (8) disk scale height(s); up to four values are given depending on the number of components (1 or 2) and their symmetry (Gaussian) or asymmetry (exponential), and (9) the FUV - NUV disk color. We correct $F_{\text{d}}$ for Galactic extinction following \citet{2007ApJS..173..293W}; this treatment assumes the extinction law of \citet{1989ApJ...345..245C} with $R_{\text{V}} = 3.1$, yielding $A_{\lambda}/E(B-V) = 8.24$ and $8.20$ in the FUV and NUV, respectively. We adopt $E(B-V)$ from \citet{2011ApJ...737..103S}. Scale heights without error bars are fixed to the value of the prior defined in \S~\ref{sec:fit_proc_indiv}. The full table is available in the online-only materials.}
\end{deluxetable*}

We show the fractional deconvolved flux in the disk at $|z| > b_{\text{25}}$, $f_{\text{d,out}}' = F_{\text{d,out}}/F_{\text{d}}$, in Fig.~\ref{fig:f_frac_nuv} (here the prime notation is used to distinguish this quantity from that which defines the denominator as the total flux, $f_{\text{d,out}} = F_{\text{d,out}}/(F_{\text{d}} + F_{\text{h}})$). The distributions are similar in both bands and are strongly peaked toward minimal disk flux outside of $b_{\text{25}}$. In the NUV (FUV), we find a median $f_{\text{d,out}}' = 0.03$ ($f_{\text{d,out}}' = 0.02$). Additionally, this is a minority of the total deconvolved flux at $|z| > b_{\text{25}}$ in galaxies with halos. Thus, in general, the use of $b_{25}$ to separate the planar and extraplanar emission is reasonable. At the same time, $35$\% of galaxies have $f_{\text{d,out}}' \geq 0.05$ in the NUV; this number is $32$\% in the FUV.

The light in the extended disk may come from several sources. \citet{2018ApJ...862...25J} fit the \textit{GALEX} FUV and NUV profiles of 38 nearby, edge-on disk galaxies with single exponential profiles, finding a modest median scale height of $h \approx 0.5$ kpc, consistent with a thick disk. They demonstrate a strong positive correlation between the UV and H$\alpha$ scale heights, suggesting that the extended UV emission comes from scattered light in the thick disk of the dusty ISM. However, it is also possible that the H$\alpha$-emitting gas is at least partially ionized by a thick stellar disk of hot, low-mass evolved stars \citep{2011MNRAS.415.2182F}, which may be responsible for producing some of the UV emission. Additionally, due to the large sample size, we adopt a simple definition of the region considered to be extraplanar, and there are instances in which perturbations to the structure of the stellar disk (e.g., warps) produce emission in this region. The disk wings at $|z| > b_{\text{25}}$ are likely a combination of starlight and scattered light from a dusty disk-halo interface. We therefore consider the disk light at $|z| > b_{\text{25}}$ to be an upper limit on additional extraplanar emission, above that measured from the halo light at $|z| > b_{\text{25}}$.

\begin{deluxetable*}{cccccccc}
  \tablecaption{Halo properties of the sample galaxies}
  \tablehead{
    \colhead{(1)} &
    \colhead{(2)} &
    \colhead{(3)} &
    \colhead{(4)} & 
    \colhead{(5)} &
    \colhead{(6)} &
    \colhead{(7)} \\
    \colhead{Galaxy} &
    \colhead{$F_{\text{h}}$, NUV} &
    \colhead{$F_{\text{h}}$, FUV} &
    \colhead{$f_{\text{h,out}}$, NUV} &
    \colhead{$f_{\text{h,out}}$, FUV} &
    \colhead{$h_{\text{h}}$} & 
    \colhead{FUV - NUV ($|z| > b_{\text{25}}$)} \\
    \colhead{} &
    \colhead{(mJy)} &
    \colhead{(mJy)} &
    \colhead{} &
    \colhead{} &
    \colhead{(kpc)} &
    \colhead{(mag)}
  }
  \startdata
  \noalign{\smallskip}
  NGC0085B & $0.04^{+0.02}_{-0.01}$ & $0.018^{+0.009}_{-0.005}$ & $0.06$ & $0.07$ & $4.2^{+2.6}_{-2.0}$ & $0.65^{+0.06}_{-0.10}$ \\
  IC0045 & $0.09^{+0.04}_{-0.03}$ & - & $0.17$ & - & $5.8^{+1.6}_{-2.3}$ & - \\
  PGC087843 & $0.06 \pm 0.02$ & - & $0.25$ & - & $5.5^{+1.8}_{-2.5}$ & - \\
  ... & ... & ... & ... & ... & ... & ... \\
  \label{tab:halos}
  \enddata
  \tablecomments{(1) Galaxy name, (2) total halo flux (NUV), (3) total halo flux (FUV), (4) fractional flux in the halo at $|z| > b_{\text{25}}$, $f_{\text{h,out}} = F_{\text{h,out}}/(F_{\text{d}} + F_{\text{h}})$ (NUV), (5) fractional flux in the halo at $|z| > b_{\text{25}}$ (FUV), (6) halo scale height, (7) FUV - NUV halo color at $|z| > b_{\text{25}}$. The halo flux is corrected for Galactic extinction as described in Tab.~\ref{tab:disks}. The full table is available in the online-only materials.}
\end{deluxetable*}

We now turn to the halo component. A halo is detected in $38$ of the sample galaxies, for a detection rate of $7$\%. For $23$ of these galaxies, data are available in both bands; the remainder are detected in the NUV alone. The halo properties are given in Tab.~\ref{tab:halos}, including the halo flux, the fractional halo flux at $|z| > b_{\text{25}}$, the halo scale height, and the halo color (FUV - NUV) at $|z| > b_{\text{25}}$.

\begin{figure*}
\centering
\includegraphics[scale = 0.75]{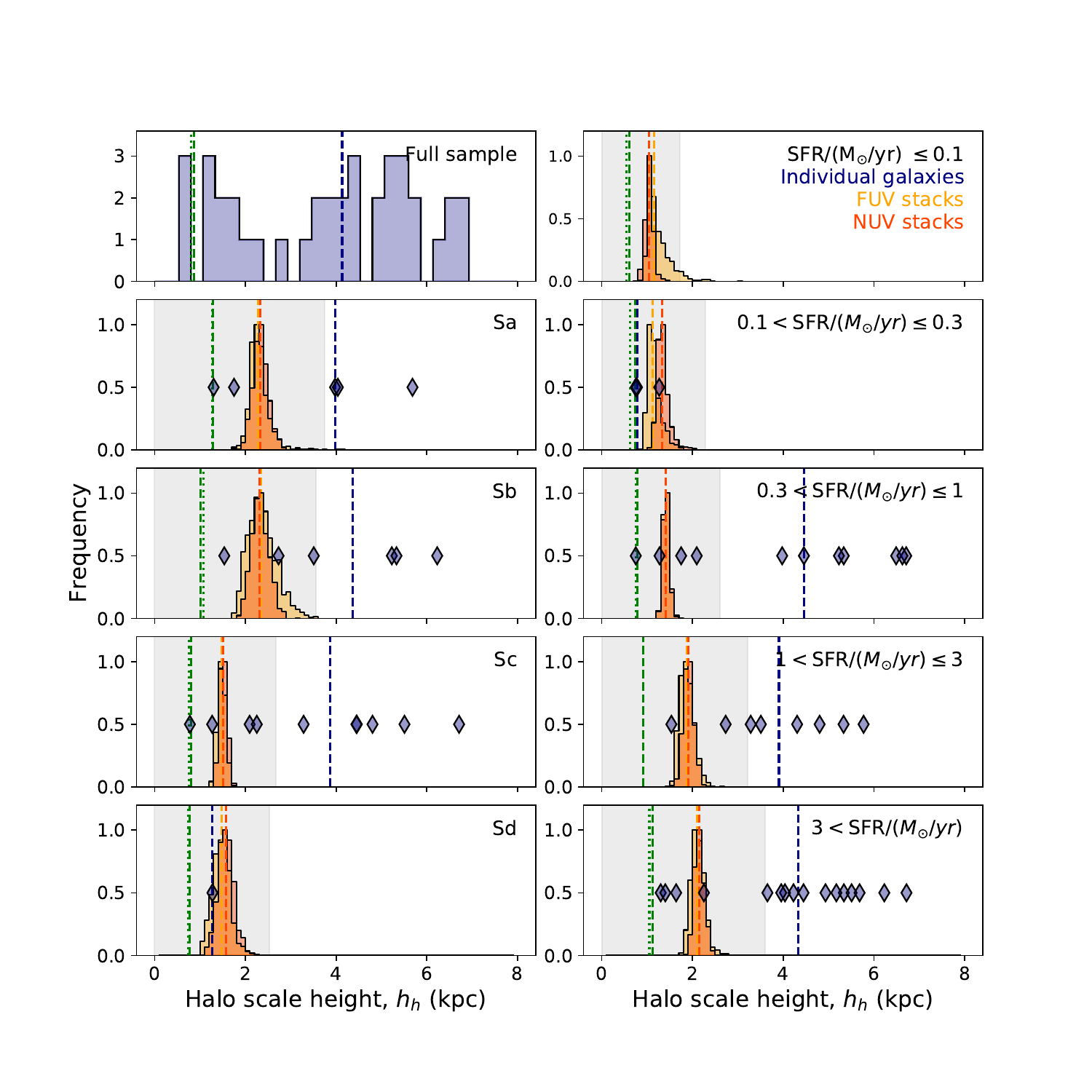}
\caption{Halo scale heights exceed the disk scale heights by factors of $\approx 4$ in the individual galaxies (blue histogram at upper left) and $\approx 2$ in the FUV (orange) and NUV (red) stacks. The blue dashed lines show the median halo scale heights, and the green dashed (NUV) and dotted (FUV) lines show the median disk scale heights. The blue diamonds indicate the halo scale heights of individual galaxies that fall into the respective morphology (left) and SFR bins (right). For the stacks, the distributions are determined from bootstrap resampling, and the gray shading shows the regions within $|z| < b_{25}$.}
\label{fig:h_halo}
\end{figure*}

We show the distribution of halo scale heights in the top left panel of Fig.~\ref{fig:h_halo}. The median halo scale height, $h_{\text{h}} = 4.1$ kpc, is a factor of $4$ larger than the median disk scale height, demonstrating that the halo is a physically distinct component. There is a significant range in $h_{\text{h}}$, with 68\% of the sample falling between $h_{\text{h}} = 1.4$ kpc and $h_{\text{h}} = 5.8$ kpc. Notably, the scale heights are resolved by the PSF and can be accurately measured. In Fig.~\ref{fig:h_halo}, we again show the distributions of halo scale heights separated by morphological classification and SFR, but the small number of galaxies in each bin preclude conclusions about trends with either property from the individual galaxies alone.

\begin{figure*}
\centering
\includegraphics[scale = 0.75]{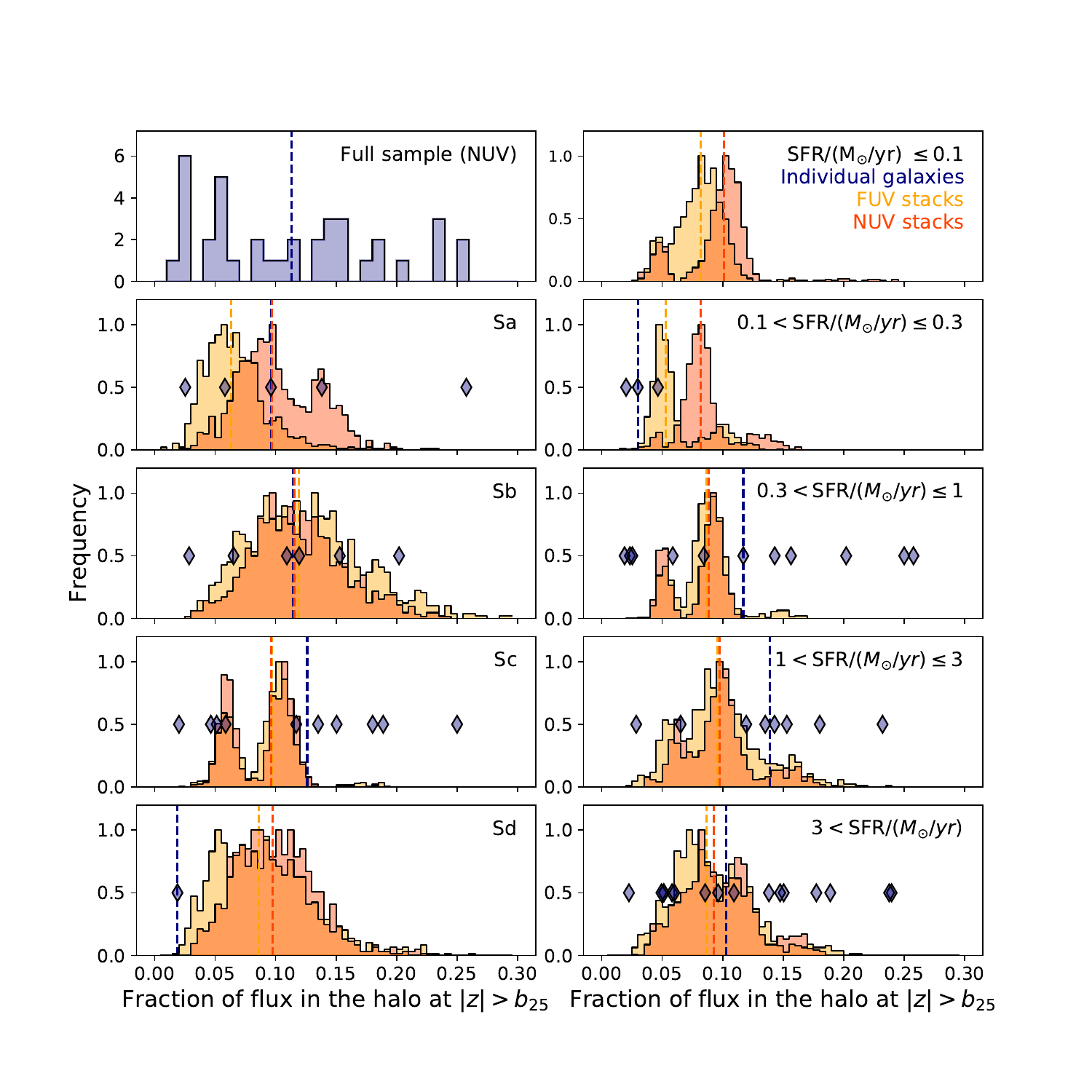}
\caption{Approximately $10$\% of the total (reddened) flux is typically found in the halo at $|z| > b_{\text{25}}$. The stacks are bootstrap resampled, and the colors and symbols are the same as in Fig.~\ref{fig:h_halo}. For the individual galaxies, we show the results from the NUV data, as $40$\% of the galaxies with detected halos only have data in this band.}
\label{fig:f_frac_halo}
\end{figure*}

In Fig.~\ref{fig:f_frac_halo}, we show the fraction of the total NUV flux found in the deconvolved halo beyond $b_{25}$, $f_{\text{h,out}} = F_{\text{h,out}}/(F_{\text{d}} + F_{\text{h}})$; this band is selected because approximately half of the sample with halo detections have NUV data alone. 68\% of the sample falls between $f_{\text{h,out}} = 0.03 - 0.2$, with a median $f_{\text{h,out}} = 0.11$. In galaxies with halo detections, the median ratio of the halo to galaxy flux at $|z| > b_{\text{25}}$ is $F_{\text{h,out}}/F_{\text{d,out}} = 7$. Combined with the significantly larger scale heights of the halos, it is clear that the halos are a distinct physical component from the disks. We defer discussion of the color of the halos to the stacking analysis in \S~\ref{sec:stack_results} due to large uncertainties on the color measurements in individual galaxies. In summary, in $\approx 10$\% of our sample, we detect extraplanar emission with a typical brightness of $\approx 10$\% of the (reddened) galaxy.

\subsection{Stacking Results}\label{sec:stack_results}

\begin{figure*}
\centering
\includegraphics[scale = 0.75]{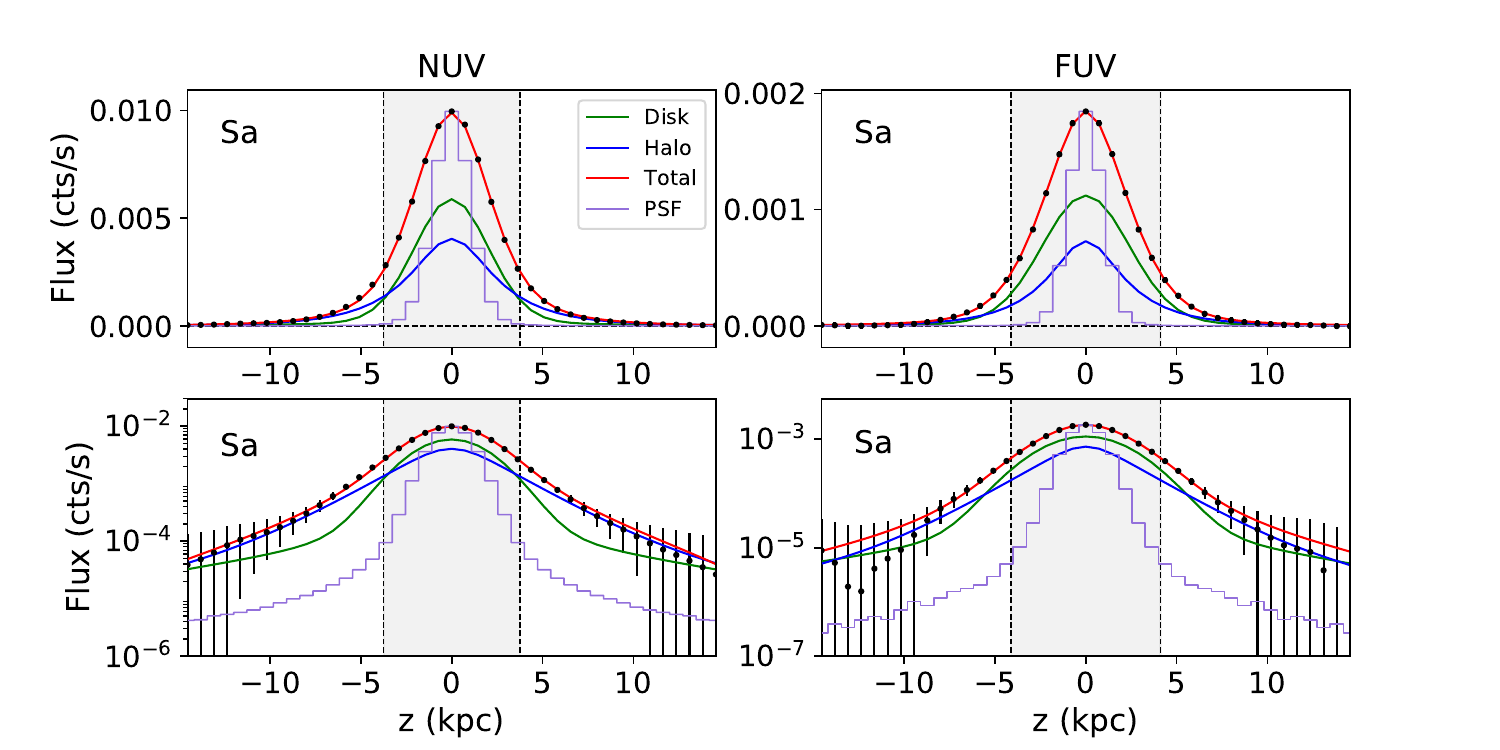}
\caption{Example stacked profiles in the NUV (left) and FUV (right) for Sa galaxies, shown on a linear (top) and log scale (bottom). The stacked data are shown by the black points; the stacks are constructed as described in \S~\ref{sec:stack_proc}, including convolution of the individual, projected profiles with the \textit{GALEX} PSF before combination at $D_{\text{max}} = 100$ Mpc. The total, PSF-convolved model is indicated in red, with the disk and halo components shown in green and blue, respectively. The PSF-convolved disk models fall below the observed profiles at $|z| < b_{\text{25}}$ at high significance, clearly demonstrating the presence of a halo. A cross-section of the \textit{GALEX} PSFs are illustrated by the purple histograms.  The shaded regions indicate where $|z| \leq b_{\text{25}}$. The profiles for all morphology and SFR stacks are shown in Appendix~\ref{sec:lum_prof}.}
\label{fig:stack_Hubble_Sd}
\end{figure*}

Next we consider evidence for extraplanar emission from the galaxy stacks. In addition to the morphology and SFR stacks, we also analyzed stacks binned by proxies for specific SFR (i.e., SFR normalized by $B$-band absolute magnitude and disk area), but we do not find significant differences compared to binning by SFR alone. In Fig.~\ref{fig:stack_Hubble_Sd}, we show an example of the stacked profiles and best-fit models for the Sa galaxies; the equivalent figures for the remaining three morphology stacks and five SFR stacks are included in Appendix~\ref{sec:lum_prof}. The results of the profile fitting are summarized in Tab.~\ref{tab:stack_results}, including the disk and halo scale heights, the fractional disk and halo fluxes at $|z| > b_{\text{25}}$, and the FUV - NUV color in the disk and in the halo at $|z| > b_{\text{25}}$.

Extraplanar emission is detected ubiquitously in the stacks. Of the $10^{3}$ bootstrap resamples that are fit with one-component (disk) and two-component (disk and halo) models, the two-component model is preferred ($p \le 0.05$) 100\% of the time for all of the stacks with the exception of the Sa galaxies in the FUV, where it is preferred $99.8$\% of the time. We consider first the properties of the disk. In Fig.~\ref{fig:hgal_nuv} and Fig.~\ref{fig:hgal_fuv}, we show the distributions of the disk scale heights, $h_{\text{d}}$. For the morphology stacks, the median NUV (FUV) $h_{\text{d}}$ decreases from $h_{\text{d}} = 1.7$ kpc ($h_{\text{d}} = 2.2$ kpc) in the Sa stack to $h_{\text{d}} = 1.1$ kpc ($h_{\text{d}} = 1.1$ kpc) in the Sd stack, consistent with the expectation that thinner disks arise in later type galaxies. The SFR stacks have $h_{\text{d}}$ values in a similar range, with a trend toward larger scale heights in galaxies with higher SFRs. The galaxies with the highest SFRs are generally the most massive systems in the sample; the median luminosity ratio between the highest and lowest SFR bins is 3.3, and the median major axis ratio is $1.6$. This is similar to the scale height ratio of $1.8$, suggesting approximate self-similarity in the disk geometry across the sample. The disk scale height distributions are tightly peaked, with typical standard deviations of $\sigma = 0.1$ kpc. 

The best-fit disk model is essential to quantify the amount of light from the stellar disk and measure the amplitude of a halo component, if any. Nevertheless, the measured disk scale heights are not expected to be physically meaningful due to the stacking procedure, in which we stack resolved disks (at least in the direction of the major axis) with different scale lengths, scale heights, fluxes, inclination angles, and potential misalignments in the midplane identification. Assuming that each disk is perfectly described by an exponential surface brightness model, stacking disks with normally distributed properties -- even at the same distance -- will introduce a Gaussian core to the stacked profile whose characteristic width does not represent any average galaxy. Furthermore, the galaxies in our sample are not perfect exponential disks: some have central dust lanes, clumps from young star clusters found off of the midplane, and/or minor warps (we select against obvious warps). Each of these features tend to broaden the best-fit disk profile. Finally, the major axis of each galaxy in the stack is resolved. Therefore, when projecting the luminosity profiles of the sample galaxies to a common (larger) distance, the projected profiles of closer galaxies have an effective PSF that is narrower than that of farther galaxies, rendering the profiles of closer galaxies ``too sharp''. While this is resolved by convolving the profile with the PSF after projection (see discussion in \S~\ref{sec:stack_proc}), this correction reduces the recoverable information about the true scale height of the projected galaxy disks.

In general, each of the effects discussed above will tend to broaden the measured disk scale heights. They will also tend to make the profiles more closely resemble a Gaussian, although we expect the wings of the best-fit profiles to still resemble an exponential disk. Indeed, the disk scale heights measured from the stacks are larger than the typical scale heights measured in the individual galaxies by up to a factor of $2 - 3$. Additionally, these factors result in elevated fractional disk fluxes at $|z| > b_{25}$. Indeed, $f_{\text{d,out}}'$ is comparable to the median $f_{\text{d,out}}' = 0.03$ measured in the individual galaxies in approximately half of the NUV stacks, while it exceeds this value by a factor of $2 - 3$ in the remaining NUV stacks. Nevertheless, as long as the light from the galaxy disk is appropriately captured by the disk model, we expect the stacks to be sensitive to extended emission on scales comparable to and exceeding the measured disk scale heights, just as stacking unresolved galaxies can reveal extended emission in other surveys (e.g., eROSITA stacks, \citealt{Zhang2024}; Lyman alpha nebula stacks, \citealt{2011ApJ...736..160S}). In summary, the disk-distorting effects discussed above all tend to be captured by broadening the bright, central disk, and the stacked galaxies clearly show additional emission beyond the best-fit disk models, as discussed next.

\begin{figure}
\includegraphics[scale = 0.35]{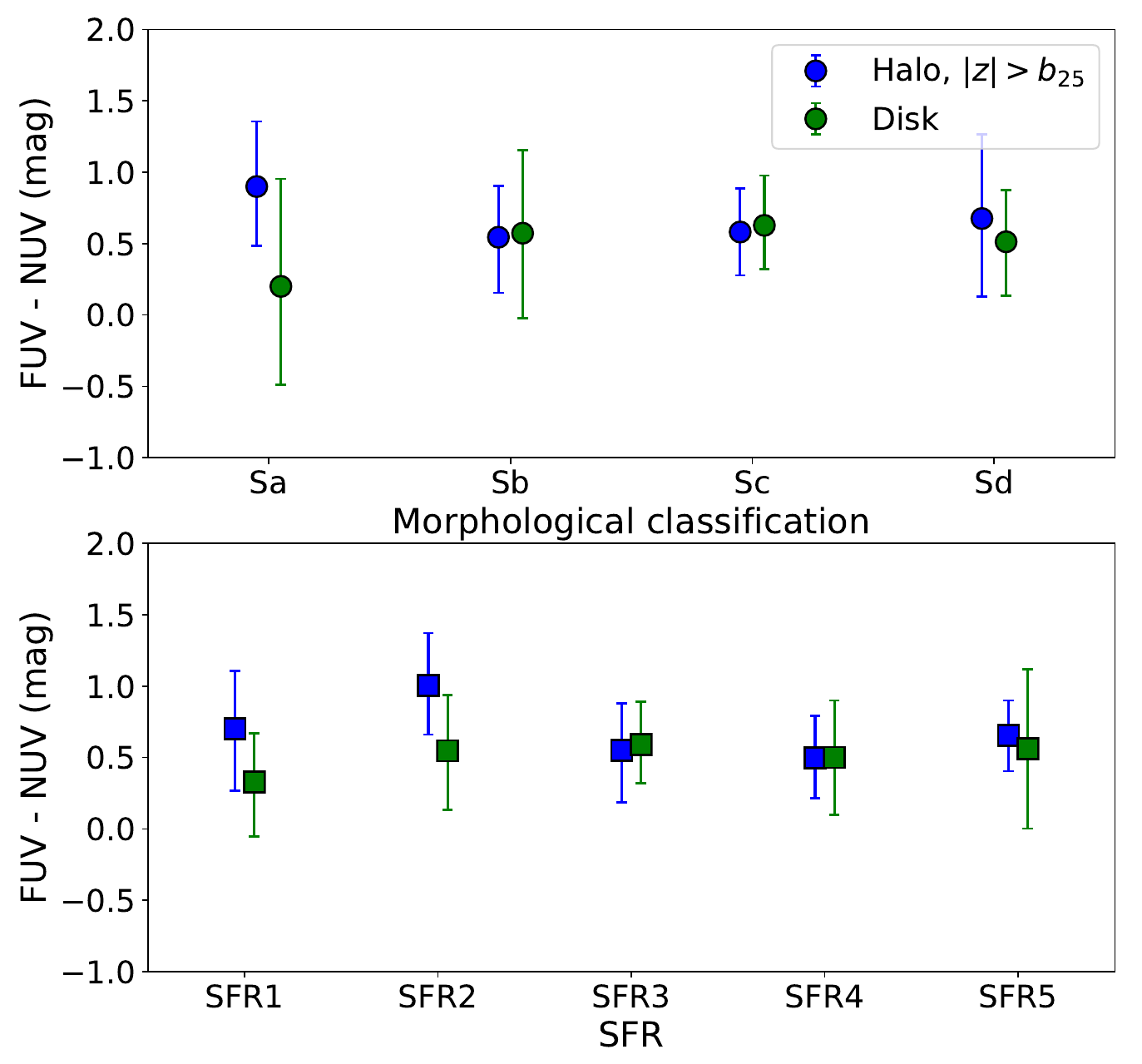}
\caption{FUV - NUV colors of the disk (green) and extraplanar emission (blue) in the morphology (top) and SFR stacks (bottom; SFR1 and SFR5 refer to the lowest and highest SFR bins, respectively). Here, the extraplanar emission is the halo flux at $|z| > b_{25}$. The data points and error bars indicate the median and 68\% confidence intervals of the bootstrapped distributions, respectively. The colors of the disks and extraplanar emission are consistent with each other within the errors, an expected result for a Milky Way-like dust composition (see \S~\ref{sec:case_for_dust}).}
\label{fig:stack_color}
\end{figure}

In Fig.~\ref{fig:h_halo}, we show the halo scale height distributions for the stacks. The median scale heights are consistent between the bands at the $\leq 1\sigma$ level. As with the disk scale heights, the halo scale heights decrease with later-type morphology and increase with higher SFR, with median values ranging from $1.1~\text{kpc} < h_{\text{h}} < 2.3~\text{kpc}$. The median $h_{\text{h}}$ measured in the individual galaxies, $h_{\text{h}} = 4.1$ kpc, falls significantly outside of this range. The larger characteristic halo scale heights measured in the individual galaxies may reflect the challenge of detecting relatively compact, faint halos under the bright disk at the typical S/N of the \textit{GALEX} data (see discussion in \S~\ref{sec:detect_rate}). At the high S/N of the stacks, however, these compact halos become detectable. In general, the halo scale height distributions are tightly peaked, with standard deviations of $\sigma = 0.1 - 0.2$ kpc.

We consider the fraction of the total flux found in the halo at $|z| > b_{25}$, $f_{\text{h,out}}$, for the stacks in Fig.~\ref{fig:f_frac_halo}. In the NUV, the median $f_{\text{h,out}}$ falls in the range $0.1 \le f_{\text{h,out}} \le 0.12$ for the morphology stacks ($0.09 \le f_{\text{h,out}} \le 0.11$ for the SFR stacks). There are not clear trends in the evolution of $f_{\text{h,out}}$ with morphology or SFR. In general, the characteristic $f_{\text{h,out}}$ in the stacks is comparable to that observed in individual galaxies ($\approx10$\%). Due to their larger scale heights, the halos are not as sensitive as the disks to the factors discussed above that produce elevated measurements of the disk scale heights and fractional fluxes at $|z| > b_{25}$.

In Fig.~\ref{fig:stack_color}, we compare the FUV - NUV color of the halo at $|z| > b_{25}$ to that of the disk. The disk colors fall in the range FUV - NUV $= 0.2 - 0.6$ mag, consistent with expectations for disk galaxies (e.g., \citealt{GildePaz2007}). In all of the stacks, the median halo colors at $|z| > b_{\text{25}}$ are comparable to or redder than the median disk colors by up to $0.7$ mags; however, the differences are not statistically significant. We discuss the interpretation of the halo colors in \S~\ref{sec:case_for_dust}. 

\begin{deluxetable*}{ccccccccc}
  \tablecaption{UV halo properties from stacking analysis}
  \tablehead{
    \colhead{(1)} &
    \colhead{(2)} &
    \colhead{(3)} &
    \colhead{(4)} & 
    \colhead{(5)} &
    \colhead{(6)} &
    \colhead{(7)} &
    \colhead{(8)} &
    \colhead{(9)} \\
    \colhead{Stack} &
    \colhead{$N_{\text{gal}}$} &
    \colhead{$\langle M_{\text{B}} \rangle$} &
    \colhead{$h_{\text{d}}$} & 
    \colhead{$h_{\text{h}}$} &
    \colhead{$f_{\text{d,out}}'$} &
    \colhead{$f_{\text{h,out}}$} &
    \colhead{FUV - NUV} &
    \colhead{FUV - NUV} \\
    \colhead{} &
    \colhead{} &
    \colhead{(mag)} &
    \colhead{(kpc)} & 
    \colhead{(kpc)} &
    \colhead{} &
    \colhead{} &
    \colhead{(disk; mag)} &
    \colhead{(extraplanar; mag)} 
  }
  \startdata
  \noalign{\smallskip}
  \multicolumn{8}{c}{\textbf{NUV}} \\
  \smallskip
  \textbf{Sa} & 31 & $-20.5$ & $1.7 \pm 0.2$ & $2.3 \pm 0.2$ & $0.03^{+0.05}_{-0.02}$ & $0.10^{+0.05}_{-0.03}$ & $0.2^{+0.8}_{-0.7}$ & $0.9^{+0.5}_{-0.4}$ \\
  \textbf{Sb} & 78 & $-20.6$ & $1.6 \pm 0.1$ & $2.3 \pm 0.2$ & $0.04^{+0.05}_{-0.03}$ & $0.12 \pm 0.04$ & $0.6 \pm 0.6$ & $0.5 \pm 0.4$ \\
  \textbf{Sc} & 225 & $-20.2$ & $1.2 \pm 0.1$ & $1.5 \pm 0.1$ & $0.06^{+0.03}_{-0.04}$ & $0.10^{+0.02}_{-0.04}$ & $0.6 \pm 0.3$ & $0.6 \pm 0.3$ \\
  \textbf{Sd} & 20 & $-19.6$ & $1.1 \pm 0.1$ & $1.6 \pm 0.2$ & $0.03^{+0.03}_{-0.02}$ & $0.10^{+0.04}_{-0.03}$ & $0.5 \pm 0.4$ & $0.7^{+0.6}_{-0.5}$ \\
  \multicolumn{8}{c}{---} \\
  \smallskip
  \textbf{$\text{SFR} \le 0.1$} & 17 & $-19.5$ & $0.9 \pm 0.1$ & $1.0 \pm 0.1$ & $0.08 \pm 0.04$ & $0.11^{+0.01}_{-0.03}$ & $0.3^{+0.3}_{-0.4}$ & $0.7 \pm 0.4$ \\
  \textbf{$0.1 < \text{SFR} \le 0.3$} & 83 & $-19.8$ & $1.1 \pm 0.1$ & $1.3 \pm 0.1$ & $0.05^{+0.04}_{-0.02}$ & $0.09^{+0.02}_{-0.01}$ & $0.5 \pm 0.4$ & $1.0^{+0.4}_{-0.3}$ \\
  \textbf{$0.3 < \text{SFR} \le 1$} & 153 & $-20.1$ & $1.17^{+0.05}_{-0.04}$ & $1.4 \pm 0.1$ & $0.06^{+0.01}_{-0.04}$ & $0.09^{+0.01}_{-0.04}$ & $0.6 \pm 0.3$ & $0.5^{+0.3}_{-0.4}$ \\
  \textbf{$1 < \text{SFR} \le 3$} & 191 & $-20.4$ & $1.4 \pm 0.1$ & $1.9 \pm 0.1$ & $0.04 \pm 0.02$ & $0.10^{+0.02}_{-0.03}$ & $0.5 \pm 0.4$ & $0.5 \pm 0.3$ \\
  \textbf{$\text{SFR} > 3$} & 87 & $-20.6$ & $1.6^{+0.1}_{-0.2}$ & $2.1 \pm 0.1$ & $0.04^{+0.05}_{-0.03}$ & $0.10 \pm 0.03$ & $0.6 \pm 0.6$ & $0.7^{+0.2}_{-0.3}$ \\
  \multicolumn{8}{c}{\textbf{FUV}} \\
  \smallskip
  \textbf{Sa} & 16 & $-20.5$ & $2.2 \pm 0.2$ & $2.3^{+0.2}_{-0.1}$ & $0.07^{+0.05}_{-0.04}$ & $0.06^{+0.03}_{-0.02}$ & & \\
  \textbf{Sb} & 48 & $-20.5$ & $1.6 \pm 0.2$ & $2.3^{+0.4}_{-0.3}$ & $0.04^{+0.06}_{-0.03}$ & $0.12^{+0.06}_{-0.05}$ & & \\
  \textbf{Sc} & 137 & $-20.2$ & $1.1 \pm 0.1$ & $1.5 \pm 0.1$ & $0.04^{+0.03}_{-0.04}$ & $0.10^{+0.02}_{-0.04}$ & & \\
  \textbf{Sd} & 14 & $-19.6$ & $1.1 \pm 0.1$ & $1.5 \pm 0.2$ & $0.04^{+0.04}_{-0.03}$ & $0.09 \pm 0.04$ & & \\
  \multicolumn{8}{c}{---} \\
  \smallskip
  \textbf{$\text{SFR} \le 0.1$} & 13 & $-19.5$ & $0.9 \pm 0.1$ & $1.1^{+0.3}_{-0.1}$ & $0.10 \pm 0.04$ & $0.09 \pm 0.02$ & & \\
  \textbf{$0.1 < \text{SFR} \le 0.3$} & 48 & $-19.7$ & $1.0 \pm 0.1$ & $1.1^{+0.2}_{-0.1}$ & $0.04^{+0.12}_{-0.02}$ & $0.05^{+0.04}_{-0.01}$ & & \\
  \textbf{$0.3 < \text{SFR} \le 1$} & 97 & $-20.0$ & $1.12^{+0.05}_{-0.04}$ & $1.4 \pm 0.1$ & $0.05^{+0.01}_{-0.04}$ & $0.09^{+0.01}_{-0.03}$ & & \\
  \textbf{$1 < \text{SFR} \le 3$} & 124 & $-20.4$ & $1.3 \pm 0.1$ & $1.9^{+0.2}_{-0.1}$ & $0.03^{+0.03}_{-0.02}$ & $0.10^{+0.03}_{-0.04}$ & & \\
  \textbf{$\text{SFR} > 3$} & 52 & $-20.6$ & $1.6^{+0.1}_{-0.2}$ & $2.1^{+0.2}_{-0.1}$ & $0.03^{+0.04}_{-0.02}$ & $0.09^{+0.04}_{-0.03}$ & & \\
  \label{tab:stack_results}
  \enddata
  \tablecomments{(1) Selection criterion for the galaxies in the stack, where the SFR has units of M$_{\odot}$/yr, (2) number of galaxies in stack, (3) average $M_{\text{B}}$ of the galaxies in the stack, weighted by $t_{\text{exp}}$, (4) scale height of Gaussian disk profile, (5) scale height of exponential halo profile, (6) fraction of the disk flux found at $|z| > b_{\text{25}}$, $f_{\text{d,out}}' = F_{\text{d,out}}/F_{\text{d}}$, (7) fraction of the total flux found in the halo at $|z| > b_{\text{25}}$, $f_{\text{h,out}} = F_{\text{h,out}}/(F_{\text{d}} + F_{\text{h}})$, (8) FUV - NUV color of the disk, and (9) FUV - NUV color of the halo at $|z| > b_{\text{25}}$.}
\end{deluxetable*}

\subsection{\textit{GALEX} PSF}\label{sec:psf}

We investigated how uncertainty in the \textit{GALEX} PSF affects the measured extraplanar emission. As shown in \citet{Morrissey2007}, the PSF changes across the FOV, primarily due to the detector. The PSF broadens and stretches with distance from the optical axis. The broadening, added in quadrature to the on-axis value, produces with a maximal increase in FWHM of about $3''$ at the field edge. The stretching also increases the ellipticity of the PSF near the field edge; both the stretching and broadening distort a galaxy's image. On the other hand, the extended wings of the PSF do not appear to change beyond the overall stretch. \citet{2016ApJ...833...58H} measured the \textit{GALEX} PSF beyond $R = 100''$ for bright sources within the inner 25\% of the FOV and measured the $R^{-\alpha}$ power law of the wings. While it is generally not possible to measure radial profiles to this distance near the edge of the FOV, we followed their method to create profiles for bright, isolated stars within the outer 25\% of the FOV and find that the power law is about the same as in the center of the FOV: $\alpha = 2$ in the FUV and $3$ in the NUV. Therefore, we anticipate that any impact of PSF variation on the measured extraplanar emission will be due to the PSF core and not to the wings.

To quantify the impact of PSF core variation, we simulated \textit{GALEX} images of exponential disks observed with a PSF with a broadened core ($6''$ and $8''$ FWHM, to capture the maximal broadening expected at the edges of the FOV). We then processed and fit these simulated galaxies using the nominal PSF, following the procedures above (\S~\ref{sec:analysis}). The broadened PSF core tends to artificially increase the scale height of the disk, but it does not create an artificial halo. We repeated this analysis for simulated galaxies with halos, assuming halo properties that are typical of the sample. The broadened PSF core again increases the disk scale height, but it does not affect our ability to reliably recover the halo scale height and luminosity, within the errors. We note that the broadened PSF core decreases the halo detection rate of faint, compact ($h_{\text{h}} \lesssim 2$ kpc) halos, which are difficult to distinguish from the broadened disk at the typical S/N of the sample. This may contribute to the low detection rate and large scale heights of extraplanar emission in the individual galaxies compared to the stacks, where higher S/N allows more sensitivity to compact extraplanar emission. The ellipticity of the PSF does not impact the stacking analysis because the galaxies in the stacks are approximately uniformly distributed across the FOV, uniformly sampling the PSF position angle. Thus, when using the standard PSF, we expect artificially large disk scale heights (see Fig.~\ref{fig:hgal_nuv}), but simulations of exponential disks and halos show that we accurately recover the extraplanar emission at sufficient S/N. Uncertainty in the PSF largely impacts the inferred \textit{disk} properties, which is further reason that we do not ascribe physical significance to the best-fit disk models. 

Finally, we verified that, for the nominal PSF, the best-fit disk models capture the expected amount of PSF-scattered light at $|z| > b_{25}$. This is especially important for the doubly convolved stacks (see \S~\ref{sec:stack_proc}), to ensure that any extraplanar emission is appropriately partitioned between the disk and halo components. We convolved images of simulated exponential disks with the PSF, measured the amount of disk emission that is found in the extraction box (see \S~\ref{sec:prof_proc}) at $|z| > b_{25}$, and compared this quantity to the  amount of disk emission at $|z| > b_{25}$ in the best-fit, \textit{PSF-convolved} disk models for the galaxies with halo detections. We found these quantities to be consistent ($4$\% of the total disk flux in the FUV and $6$\% in the NUV), demonstrating that the disk models capture the expected amount of disk emission that is redistributed into the halo by the PSF. We performed the comparable test for the stacks by projecting and doubly convolving the simulated galaxies, and again we found comparable disk emission at $|z| > b_{25}$ in the \textit{PSF-convolved} disk models as for the simulated galaxies ($\approx 12$\% of the total disk flux in the FUV and $15$\% in the NUV). We are therefore confident that our model fitting correctly assigns the disk and halo light to their respective components in both the individual galaxies and the stacks, given the nominal PSF. If instead the PSF core is broadened, as discussed above, the disk scale height is artificially broadened, but we still reliably recover the luminosity and scale height of the extraplanar emission. Therefore, the reported properties of the extraplanar emission are robust against variations in the PSF.

\section{Discussion}
\label{sec:disc}

We have shown that extraplanar UV emission is commonplace in nearby, normal disk galaxies in a sample more than an order of magnitude larger than initial studies \citep[e.g.,][]{2014ApJ...789..131H, 2018ApJ...862...25J}. Here, we first demonstrate that the low detection fraction of extraplanar emission in the individual galaxies ($38/551$, or $7$\%) is consistent with its ubiquitous detection in the stacks, given the sensitivity limits of the data (\S~\ref{sec:detect_rate}). We then discuss the physical origin of the extraplanar emission and make the case that it arises from reflection nebulae produced by extraplanar dust (\S~\ref{sec:case_for_dust}). Finally, in \S~\ref{sec:dh_connect}, we explore implications for the circulation of gas between the disk and the halo, including the characteristic dust-bearing gas mass at the disk-halo interface and the fraction of recycled material in the inner CGM.

\subsection{Incidence of extraplanar emission}
\label{sec:detect_rate}

The low detection rate of extraplanar emission in the individual galaxies is consistent with its ubiquitous detection in the stacked galaxies at the typical depth of the \textit{GALEX} data. As discussed in \S~\ref{sec:sample}, we selected galaxies with $t_{\text{exp}} \geq 1000$ s; this exposure time was estimated to detect a $M_{\text{NUV}} \approx -13.7$ mag halo at $(4 - 5)\sigma$ significance at a distance of $D = 100$ Mpc. However, this estimate applies to the detection of the halo component alone, and does not account for the detection of faint extraplanar emission superimposed on a bright, extended disk. The challenge of separating the halo from the dominant disk, particularly when the halo is relatively compact, significantly reduces the detection fraction.

We use a simple simulation to quantify the expected detection fraction for a range of halo scale heights and exposure times. We begin with a two-dimensional model of an exponential disk and halo of the form $f(z,R) = f_{\text{0,d}}e^{-|R|/R_0}e^{-|z|/h_{\text{z,d}}} + f_{\text{0,h}}e^{-|R|/R_0}e^{-|z|/h_{\text{z,h}}}$. We assume an amplitude ratio of $f_{\text{0,h}} = 0.1f_{\text{0,d}}$, an exponential scale length of $R_0 = 3$ kpc for both components, a disk scale height of $h_{\text{z,d}} = 0.5$ kpc, and halo scale heights of $h_{\text{d}} = 1 - 6$ kpc. In our sample, the halo models typically contain $20\% - 40$\% of the total flux in the galaxy (note that this percentage includes all of the flux in the halo component, not only the flux at $|z| > b_{25}$; the majority of this flux is found at $|z| < b_{25}$ and thus is assigned to the stellar disk). We adopt the lower end of this range to assume $M_{\text{NUV,h}} = -14$ mag. We simulate the expected image for $t_{\text{exp}} = 1500 - 15000$ s at a distance of $D = 50$ Mpc, introducing Poisson deviates and including the typical background. From the two-dimensional model, we construct a one-dimensional profile as performed for the data. For 100 trials at each point in parameter space, we fit the profile with one- (disk) and two-component (disk and halo) models and assess the need for the halo using the F-test, as above.

The simulated detection rate varies significantly depending on the halo scale height, ranging from a few percent for $h_{\text{h}} = 1 - 2$ kpc to $50\% - 100$\% for $h_{\text{h}} \ge 4$ kpc, with higher detection rates found for longer exposure times. This simple test is broadly consistent with the $7$\% detection rate observed in a sample with a median $h_{\text{h}} \approx 4$ kpc and $t_{\text{exp,NUV}} = 1700$ s. This demonstrates the challenge of detecting relatively compact halos beneath the signal of (even quite compact) disks at the typical exposure times of the MIS given the \textit{GALEX} PSF. Thus, we are sensitive only to the "tip of the iceberg" of the apparent brightness distribution of the halos and, consequently, of the host galaxies given the characteristic observed $f_{\text{h,out}} \approx 0.1$. In contrast, the stacks contain several tens to $\gtrsim 200$ galaxies each, resulting in typical increases in the S/N of factors of $5 - 15$ compared to the individual galaxies. The ubiquitous detection of the halos in the stacks is thus consistent with the finding that the simulated detection fraction approaches $100$\% for exposure times an order of magnitude greater than the median $t_{\text{exp,NUV}} = 1700$ s. 

\begin{figure}
\includegraphics[scale = 0.5]{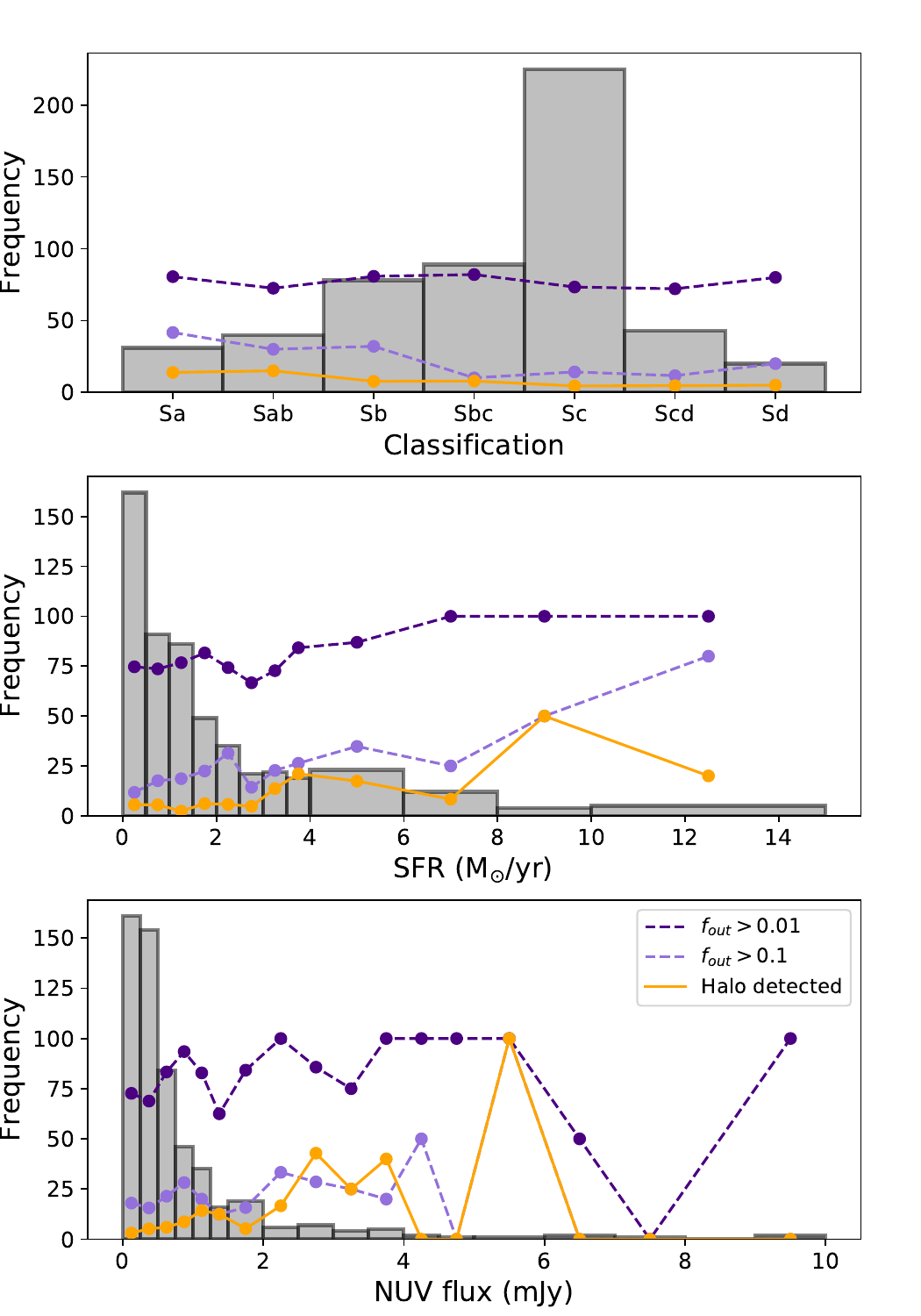}
\caption{The halo detection fractions (orange lines) are highest in galaxies with earlier morphological types (top), higher SFRs (middle), and higher total NUV fluxes (bottom). Also shown are the detection fractions of extraplanar emission at $|z| > b_{\text{25}}$ that are $>1$\% (dashed dark purple line) and $>10$\% (dashed light purple line) of the total flux, where we consider contributions to the flux outside $b_{25}$ from both the disk and the halo components. All detection fractions are multiplied by 100 for display purposes.}
\label{fig:detect_frac}
\end{figure}

In Fig.~\ref{fig:detect_frac}, we show the detection rate of extraplanar emission as a function of morphology, SFR, and total NUV flux. We also include the detection rates of total (disk and halo) flux at $|z| > b_{25}$ that exceeds 1\% and 10\% of the total flux. As shown in the bottom panel, all three detection metrics increase with the total NUV flux of the host galaxy. The halo detection fraction rises from a few percent at $F_{\text{NUV}} \lesssim 1$ mJy to $\gtrsim 25$\% at $F_{\text{NUV}} > 2$ mJy until small number statistics prevent robust trends from being determined at the highest values of $F_{\text{NUV}}$. Thus, the halo detection rate is positively correlated with the apparent brightness of the host galaxy; this result is expected given the relatively constant $f_{\text{h,out}} \approx 0.1$, suggesting that the apparently brightest halos are found in the apparently brightest galaxies.

Additionally, as shown in the top and middle panels of Fig.~\ref{fig:detect_frac}, the highest halo detection rates also occur in the galaxies with the highest SFRs and earlier morphological classifications (Sa/Sb). These galaxies tend to be the most massive and intrinsically brightest galaxies, which are likely to host the intrinsically brightest halos. We also detect halos in intrinsically fainter, later-type, and lower-SFR galaxies, preferentially among nearer galaxies in the sample. However, it is clear that the average properties of the galaxy subsample with halo detections are not representative of the full sample, and we rely on the stacking results to expand the dynamic range in morphology and SFR over which we are sensitive to extraplanar emission.

\subsection{The case for dusty halos}
\label{sec:case_for_dust}

The properties of the extraplanar emission are consistent with reflection nebulae produced by extraplanar dust (hereafter the RN model), in contrast to stellar halos (line emission produced by gaseous halos is inconsistent with the SEDs observed in previous studies of smaller samples, e.g., \citealt{2014ApJ...789..131H, 2016ApJ...833...58H}.) We can rule out stellar halos as the primary origin as follows. Using the Kinney and Calzetti spectral galaxy templates \citep{1994ApJ...429..582C, 1996ApJ...467...38K}, we adopt their Sc template for the disk and their elliptical (E) and S0 templates for an older, redder stellar halo. In this model, requiring $f_{\text{h,out}} \approx 0.1$ results in an optical halo luminosity that exceeds the disk luminosity by a factor of a few. In contrast, the bolometric luminosities of the stellar halos of nearby, massive galaxies are typically a few to $10$\% of the disk luminosity \citep[e.g.,][]{2011ApJ...736...24B, 2019MNRAS.490.3426D}. Additionally, in general, the FUV - NUV colors of evolved stellar populations fall in the range FUV - NUV = $0.5 - 2.5$ mags \citep{GildePaz2007}, with a characteristic color of FUV - NUV $\approx 1.5$ that is 0.5 - 1.0 mags redder than the median halo colors derived from the stacks (see Fig.~\ref{fig:stack_color}). In summary, the optical and UV SEDs of the extraplanar emission are inconsistent with a primary origin in an evolved stellar halo. The morphology of the extraplanar emission is also inconsistent with a primary origin in warped stellar disks. While warps occur in $\gtrsim 50$\% of nearby disk galaxies, they typically affect the outer disk ($R > r_{25}$; \citealt{Briggs1990}), thus falling outside of the extraction box. Even if warps occur in the inner disk, the typical warp angle ($\theta \approx 5^{\circ}$; \citealt{Zee2022}) is insufficient to produce projected emission at a characteristic scale height of $h_{\text{h}} \gtrsim 1$ kpc within $R < r_{25}$.

There is also positive evidence in favor of the RN model. Monte Carlo radiative transfer models generally reproduce the observed $f_{\text{h,out}}$ for reasonable SFRs and extraplanar dust masses (\citealt{2014ApJ...785L..18S, 2015ApJ...815..133S}; although not every galaxy has extraplanar emission in the UV, e.g., \citealt{2015ApJ...815..133S}). The FUV - NUV colors of the extraplanar emission in the stacks are also consistent with the RN model, assuming Milky Way-like dust. As shown in Fig.~\ref{fig:stack_color}, we observe comparable FUV - NUV colors in the disks and in the halos at  $|z| > b_{\text{25}}$ within the errors. A trend toward somewhat redder halos has been previously observed in smaller samples \citep{2016ApJ...833...58H}. In the RN model, the halo color depends on the incident spectrum and the dust grain properties. The extraplanar luminosity, $L_{\text{h}}$, is related to the disk luminosity, $L_{\text{g}}$, following $L_{\text{h}}(\lambda) = L_{\text{g}}(\lambda)(1 - \text{e}^{-\tau(\lambda)\varpi(\lambda)})$. The optical depth, $\tau(\lambda) = N_{\text{D}}\sigma_{\text{ex}}$, depends on the dust column density, $N_{\text{D}}$, and extinction cross-section, $\sigma_{\text{ex}}$, while the scattering albedo, $\varpi(\lambda) = \sigma_{\text{scat}}/\sigma_{\text{ex}}$, also depends on the scattering cross-section, $\sigma_{\text{scat}}$. Assuming an optically thin halo, the single-scattering approximation gives $L_{\text{h}}(\lambda)/L_{\text{g}}(\lambda) \approx \tau(\lambda)\varpi(\lambda) = N_{\text{D}}\sigma_{\text{scat}}(\lambda)$.

Adopting the dust models of \citet{2001ApJ...548..296W}, a Milky Way-like dust composition with $R_{V} = 3.1$ has roughly comparable extinction cross-section at the midpoint of the FUV and NUV filters, due to the 2175 \AA\ bump that affects the NUV. Thus, it is reasonable to approximate the color of the incident spectrum by the observed disk color. Additionally, the scattering cross-section at the midpoint of the FUV filter is slightly lower than at the midpoint of the NUV filter ($\sigma_{\text{scat}}$(FUV)/$\sigma_{\text{scat}}$(NUV) $\approx 0.8$). This suggests that, if the average dust properties are similar to those in the Milky Way, we expect a trend toward slightly redder extraplanar emission in the RN model. This trend is not expected for all dust types; for SMC-type dust, for example, the higher extinction cross-section and comparable scattering cross-section in the FUV compared to the NUV suggest halos that are comparable to or bluer than the disks \citep{2001ApJ...548..296W}. Supplementing the \textit{GALEX} UV spectral energy distribution with \textit{Swift} observations will provide further constraints on the typical grain properties in CGD \citep{2014ApJ...789..131H, 2016ApJ...833...58H}. In summary, the RN model is consistent with the observed properties of the extraplanar emission within the errors and provides the most physically plausible explanation for the origin of the extraplanar light.

\subsection{Implications for the disk-halo connection}
\label{sec:dh_connect}

Adopting the RN model, the ubiquitous detection of extraplanar emission in the stacks demonstrates that the circulation of dust-bearing gas on kpc scales is commonplace in galaxies with diverse morphologies and SFRs. Here we estimate the dust-bearing gas mass involved in this circulation. We can express the column density of dust-bearing gas in the halo, $N_{\text{H}}$, in terms of the ratio of the halo luminosity at $|z| > b_{\text{25}}$, $L_{\text{h,out}}$, to the total luminosity within $|z| \leq b_{\text{25}}$, $L_{\text{in}}$, or $l_{\text{out}} = L_{\text{h,out}}/L_{\text{in}}$, as $N_{\text{H}} = l_{\text{out}}(\lambda)/\sigma_{\text{scat,H}}(\lambda)$. Here, $l_{\text{out}}$ is calculated in the NUV and is corrected for reddening as described in Appendix \ref{sec:ext_corr}. The scattering cross-section is now expressed per H nucleon.

\begin{figure*}
\centering
\includegraphics[scale = 0.55]{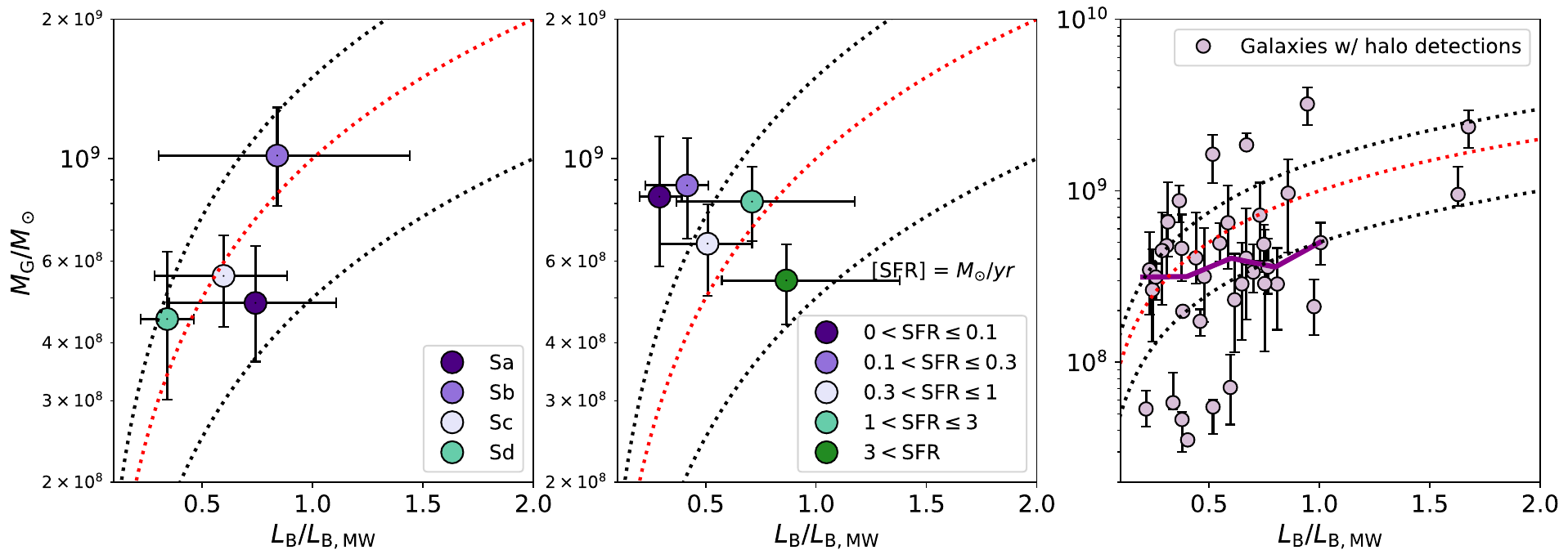}
\caption{Extraplanar dust-bearing gas mass, $M_{\text{G}}$, is typically $\approx 5 - 15$\% of the ISM mass of the host galaxy. $M_{\text{G}}$ is shown as a function of normalized $B$-band luminosity, $L_{\text{B}}/L_{\text{B,MW}}$, where $L_{\text{B,MW}}$ is the luminosity of the Milky Way. At left and center, the data points are determined from the bootstrapped morphology and SFR stacks, respectively, and the error bars encompass 68\% of the distribution. In all panels, the dashed lines show 5\% (bottom, black), 10\% (middle, red), and 15\% (top, black) of the ISM gas mass assuming the Galactic ISM mass fraction. In the individual galaxies shown at right, the median $M_{\text{G}}$ indicated by the thick magenta line is consistent with that measured from the stacks within the errors.}
\label{fig:dustbearing_mass}
\end{figure*}

The total dust-bearing gas mass is given by $M_{\text{G}} = 2\pi r_{25}^{2}N_{\text{H}}\overline{m}$, where $\overline{m} = 1.4m_{p}$ is the average mass per nucleon. Here we assume a cylindrical geometry for the dust-bearing gas layer surrounding a disk of radius $r_{25}$. For the Milky Way dust models of \citet{2001ApJ...548..296W} ($R_{V} = 3.1$, with an assumed dust-to-gas mass ratio of $\delta \sim 1/100$), we adopt $\sigma_{\text{scat,H}}$(NUV) $= 6.5 \times 10^{-22}$ cm$^{2}$/H. In the right panel of Fig.~\ref{fig:dustbearing_mass}, we show $M_{\text{G}}$ as a function of galaxy $B$-band luminosity for galaxies with halo detections. The median $M_{\text{G}}$ ranges from $M_{\text{G}} = 3 - 5~\times 10^{8}$ M$_{\odot}$ for galaxies with $L_{\text{B}}/L_{\text{B,MW}} \leq 1$, where there are sufficient sample statistics (here $L_{\text{B,MW}}$ refers to the $B$-band luminosity of the Milky Way, $M_{\text{B,MW}} = -20.8$ mag). This corresponds to an extraplanar dust mass of a few $\times ~10^{6}$ M$_{\odot}$. We compare the dust-bearing gas masses to the total ISM mass, adopting a Milky Way ISM mass of $10^{10}$ M$_{\odot}$ and assuming that the ISM mass scales proportionally with the luminosity. As shown in Fig.~\ref{fig:dustbearing_mass}, the median $M_{\text{G}}$ falls between $5\% - 15$\% of the ISM gas mass. In the left and center panels of Fig.~\ref{fig:dustbearing_mass}, we show that $M_{\text{G}}$ derived from the morphology and SFR stacks is also consistent with $5\% - 15$\% of the ISM mass within the errors. Due to the $50$\% systematic uncertainty on the reddening correction discussed in Appendix \ref{sec:ext_corr}, no trends in $M_{\text{G}}$ can be inferred with morphology or SFR.

There is significant scatter in the inferred $M_{\text{G}}$ at fixed $L_{\text{B}}$ in the individual galaxies. At the upper end of the distribution, three galaxies with $L_{\text{B}}/L_{\text{B,MW}} < 1$ have $M_{\text{G}} > 10^{9}$ M$_{\odot}$. If their measured extraplanar luminosities indeed arise from reflection nebulae, then $\approx 30$\% of their ISM gas mass is extraplanar; this is consistent with the upper end of the expected range of extraplanar cold gas masses determined by HI 21-cm observations \citep[e.g.,][]{2007AJ....134.1019O}. However, systematic uncertainties may contribute significantly to the scatter in the inferred $M_{\text{G}}$. At high inferred $M_{\text{G}}$, the measured extraplanar luminosities may be contaminated by stellar light due to morphological perturbations in the disk. Conversely, at low $M_{\text{G}}$, we may underestimate the extraplanar luminosity if extraplanar emission that is modeled in the wings of the disk in fact arises from scattered light in the halo. Additionally, variations in the dust grain properties and the dust-to-gas mass ratios may also produce scatter in Fig.~\ref{fig:dustbearing_mass}. Due to these systematics, we consider the median measured $M_{\text{G}}$ to be the most robust estimate of the characteristic extraplanar dust-bearing gas mass. Galaxy-to-galaxy variation in this quantity remains an important open question that next-generation observatories with improved sensitivity and angular resolution will address.

This study provides an independent confirmation of previous measurements of the dust mass at the disk-halo interfaces of nearby galaxies, which range from a few percent to a few tens of percent of the total dust mass. \citet{1997AJ....114.2463H} analyze absorption from extraplanar, dust-bearing gas structures in NGC 891 to conclude that at least $2$\% of the neutral ISM gas mass is found in these features. Direct detection of dust grains via their thermal emission typically suggests that at least $10\% - 15$\% of the total dust in nearby, star-forming galaxies is found outside of their disks \citep[e.g.,][]{2015ApJ...804...46M, 2021MNRAS.502..969Y}; in vigorously star-bursting systems, this number may be significantly higher \citep{2010A&A...514A..14K}. Furthermore, HI 21-cm observations indicate that a few percent to a third of the cold gas mass in nearby disk galaxies is extraplanar \citep[e.g.,][]{2019A&A...631A..50M}; this is broadly consistent with our finding that a characteristic $5\% - 15$\% of the ISM mass is found in dust-bearing gas at the disk-halo interface. The extraplanar dust detected by \textit{GALEX} may thus be found predominantly in the cold phase, with no evidence that significant partitioning with the hot phase is necessary.

We compare this gas mass to that found in the hot, volume-filling phase of the inner CGM ($R < 10$ kpc). For a Milky-Way mass galaxy, we adopt the $\beta$-profile model of the Galactic CGM from \citet{2015ApJ...815...77S} to find a hot gas mass at $R < 10$ kpc of $M_{\text{10}} \approx 8 \times 10^8$ M$_{\odot}$. This is roughly comparable to the $M_{\text{G}} \approx 5 \times 10^{8}$ M$_{\odot}$ found for Milky-Way mass galaxies in our sample, suggesting that at least a third of the inner CGM is composed of recycled material. If feedback processes have evacuated gas from the hot halo in these inner regions, then this recycled fraction is even higher. Although there is significant uncertainty in the CGM mass profile of the Milky Way, the model of \citet{2015ApJ...815...77S} provides an inner CGM mass at the upper end of current estimates (see discussion in \citealt{2018ApJ...863...49B}), and thus adoption of different models will generally increase the estimated recycled fraction at $R < 10$ kpc.

\begin{figure*}
\centering
\includegraphics[scale = 0.55]{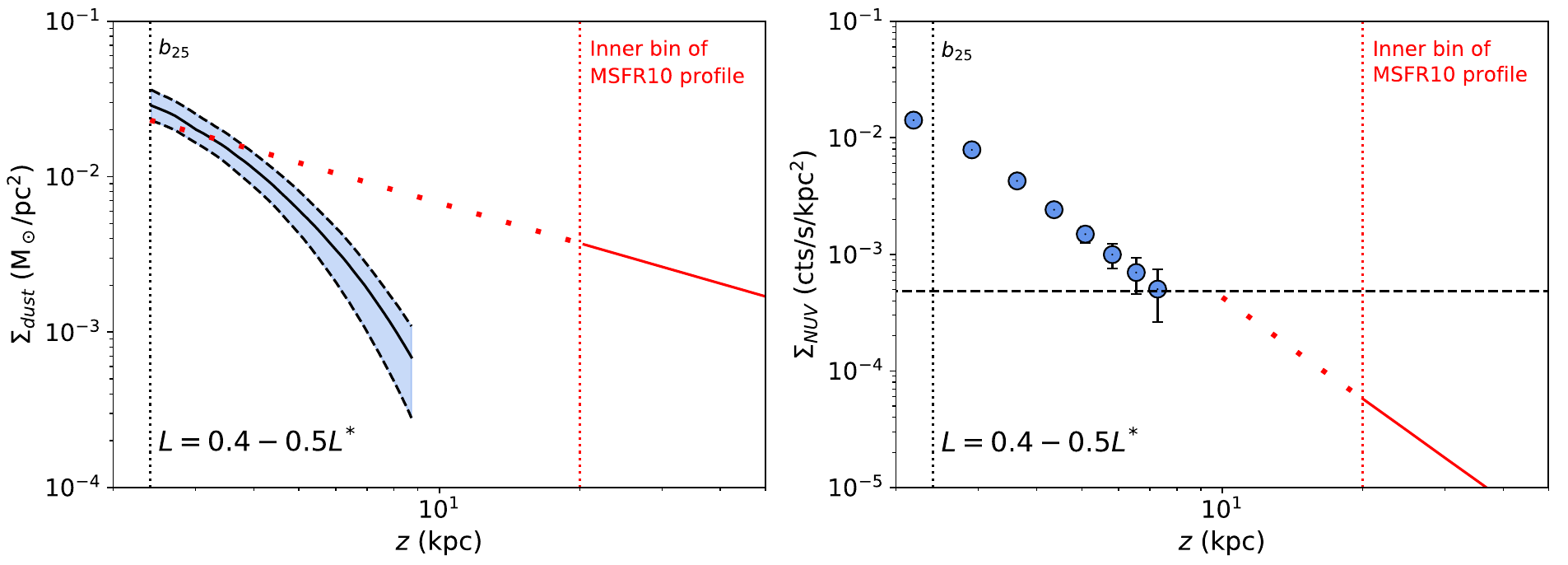}
\caption{Comparison of CGD properties at $|z| < 10$ kpc (detected in this study via reflection nebulae, in blue) with those at $R \geq 20$ kpc (derived by \citetalias{2010MNRAS.405.1025M} via reddening of background quasars, in red). We show the dust surface density, $\Sigma_{\text{dust}}$ (in blue at left) derived from the NUV surface brightness profile, $\Sigma_{\text{NUV}}$ (in blue at right), of a \textit{GALEX} stack selected to match the luminosity of the galaxies that dominate the \citetalias{2010MNRAS.405.1025M} signal ($L = 0.4-0.5L^{*}$). At left, $\Sigma_{\text{dust}}$ reported by \citetalias{2010MNRAS.405.1025M} at $R \gtrsim 20$ kpc, and its extrapolation to smaller $R$, are shown by the solid and dotted red lines, respectively. At right, we show the $\Sigma_{\text{NUV}}$ of the reflection nebula produced by illuminating the dust distribution of \citetalias{2010MNRAS.405.1025M} with a $L \approx 0.5L^{*}$ host galaxy. In the innermost spatial bin of \citetalias{2010MNRAS.405.1025M}, $R \approx 20$ kpc, $\Sigma_{\text{NUV}}$ falls almost an order of magnitude below the 2$\sigma$ background of the \textit{GALEX} stack, consistent with our non-detection of UV flux on these scales. If the \citetalias{2010MNRAS.405.1025M} profile extends smoothly to smaller $R$, the expected $\Sigma_{\text{NUV}}$ would become detectable at $R \approx 10$ kpc. The absence of a detection on this scale suggests that the dust, if present, is clumpy.} 
\label{fig:Menard_comp}
\end{figure*}

\defcitealias{2010MNRAS.405.1025M}{MSFR10}
We can ask if and how the distribution of dust at the disk-halo interface is related to that in the CGM and IGM. \citet{2010MNRAS.405.1025M} (hereafter \citetalias{2010MNRAS.405.1025M}) used reddening of background quasars at 20 kpc $-$ 10 Mpc scales around $z \approx 0.3$ galaxies in the Sloan Digital Sky Survey to probe the amount and extent of dust in the CGM and beyond. In Fig.~\ref{fig:Menard_comp}, we compare the dust surface density profiles, $\Sigma_{\text{dust}}$, from \citetalias{2010MNRAS.405.1025M} and along the minor axis of a \textit{GALEX} stack with $L_{\text{B}} = 0.4-0.5L_{\text{B}}^{*}$ (this luminosity range is selected to match the galaxies that contribute most significantly to the \citetalias{2010MNRAS.405.1025M} reddening signal). It is clear that the two detection methods probe different dust distributions, suggesting distinct physical origins. The dust surface density probed by \textit{GALEX} drops by almost two orders of magnitude and falls below the detection threshold within 10 kpc of the disk, consistent with a galactic fountain flow. Notably, extrapolation of this distribution to larger $R$ falls two orders of magnitude short in explaining the reddening signal detected in the innermost spatial bin of \citetalias{2010MNRAS.405.1025M} at $R = 20$ kpc. If CGD is found on these scales, more energetic mechanisms such as outflows or interactions are needed to transport the dust to these distances compared to the relatively quiescent fountain flows that produce the reflection nebulae detected here.

In the right panel of Fig.~\ref{fig:Menard_comp}, we ask whether the dust distribution inferred by \citetalias{2010MNRAS.405.1025M} should produce a reflection nebula that is detectable in the \textit{GALEX} data. The red line shows the NUV surface brightness, $\Sigma_{\text{NUV}}$, produced by the spherical dust distribution of \citetalias{2010MNRAS.405.1025M} when illuminated by a $L \approx 0.5L^{*}$ galaxy, assuming Milky Way-type dust. Consistent with our non-detection on $R = 20$ kpc scales, we find that $\Sigma_{\text{NUV}}$ falls almost an order of magnitude below the $2\sigma$ background of the \textit{GALEX} stack in the innermost bin of \citetalias{2010MNRAS.405.1025M}. Therefore, the UV stacking analysis cannot confirm or rule out the presence of the \citetalias{2010MNRAS.405.1025M} dust distribution at $R \gtrsim 20$ kpc, and thus we cannot assess the potential role of extended gaseous disks in enhancing the reddening signal on these scales \citep{2016MNRAS.462..331S}.

It is clear from Fig.~\ref{fig:Menard_comp} that if the $\Sigma_{\text{dust}} \propto R^{-0.86}$ profile of \citetalias{2010MNRAS.405.1025M} extends inwards to small $R$, the scattered light signal should become detectable on $R \approx 10$ kpc scales. The lack of UV detection on these scales suggests either that 1) the dust profile evolves as it approaches the disk, or 2) the dust is found in clumps and thus eludes detection via scattered light. A clumpy dust distribution is consistent with the findings of \citet{Menard2012} that a significant fraction of CGD is found in \ion{Mg}{2} absorbers. Deeper observations with next-generation UV telescopes will directly test the conclusions of \citetalias{2010MNRAS.405.1025M} at $R \approx 10 - 20$ kpc via reflection nebula measurements. 

\section{Summary and Conclusions}

We have conducted the largest search for extraplanar UV emission around nearby disk galaxies to date by more than an order of magnitude, demonstrating their ubiquitous presence in stacks of galaxies with diverse morphologies and SFRs. There is strong evidence that the extraplanar emission is reflection nebulae produced by extraplanar dust, providing a powerful probe of dust-bearing gas in the inner halos of star-forming galaxies. We summarize our key results below:

\begin{itemize}
    \item We detect extraplanar emission in $7$\% (38/551) of the individual galaxies in the \textit{GALEX} FUV and/or NUV bands, with a median scale height of $h_{\text{h}} = 4.1$ kpc and fractional (reddened) flux at $|z| > b_{25}$ of $f_{\text{h,out}} = 0.11$, where $|z| = b_{25}$ is defined as the height above the disk at which $\mu_{\text{B}} > 25$ mag arcsec$^{-2}$. There is significant diversity in the halo properties, with scale heights as large as $h_{\text{h}} = 7$ kpc and fractional fluxes as high as $f_{\text{h,out}} = 0.26$.
    \item We construct galaxy stacks based on morphology and SFR and detect extraplanar emission in all of these deep composite profiles, with scale heights in the range $h_{\text{h}} = 1.0 - 2.3$ kpc and a characteristic $f_{\text{h,out}} \approx 0.1$. With $\approx 10 - 200$ galaxies per stack, the increased S/N provides sensitivity to extraplanar emission that is fainter and/or has smaller scale heights compared to the individual galaxies. The ubiquity of the detections suggests that extraplanar UV emission is found around most normal disk galaxies at low redshift.
    \item The extraplanar emission is inconsistent with arising from a stellar halo; in order to produce the UV emission with an older, redder stellar population, an unphysically large optical luminosity is required. The characteristic extraplanar FUV - NUV color, FUV - NUV $ = 0.5 - 1$ mag, is also somewhat bluer than the typical color of an evolved stellar population (FUV - NUV $\approx 1.5$ mag; \citealt{GildePaz2007}). In contrast, the extraplanar luminosities and colors are consistent with scattered starlight from Milky Way-like dust.
    \item In the reflection nebula model, the extraplanar flux fractions typically yield a dust-bearing gas mass of $5\% - 15$\% of the ISM gas mass. This represents a significant reservoir of enriched baryons transported from the disk to the inner halo ($M_{\text{G}} \approx 5 \times 10^{8}$ M$_{\odot}$ in a Milky Way mass galaxy). The implied recycled fraction of the CGM at $R < 10$ kpc is at least one third.
    \item Comparison of the dust surface density profiles derived in this work at $|z| < 10$ kpc with that determined by \citetalias{2010MNRAS.405.1025M} from quasar reddening at $R = 20$ kpc $-$ 10 Mpc demonstrates that these dust distributions arise from different physical processes. The scale heights and gas masses determined from the reflection nebulae are consistent with a galactic fountain flow, while the dust distribution inferred by \citetalias{2010MNRAS.405.1025M} necessitates episodic galactic winds and/or stripping in interactions to transport the dust to CGM scales.
\end{itemize}

This statistically significant demonstration of the commonality of a large reservoir of dust-bearing gas at the disk-halo interfaces of nearby, star-forming galaxies motivates further study of extraplanar dust with next-generation UV telescopes. The Ultraviolet Explorer (UVEX; \citealt{2021arXiv211115608K}) will improve on the sensitivity of \textit{GALEX} by a factor of 50 at more than twice the angular resolution, enabling detections of extraplanar emission in most nearby, massive galaxies in the 2030s. These observations will allow more precise measurements of the extraplanar dust-bearing gas mass and permit spatial variations in the nebula properties to be probed. Such observations will improve constraints on the physical mechanisms responsible for elevating the dust-bearing gas into the lower halo and driving the lifecycle of dust grains within gaseous galactic ecosystems.   

\begin{acknowledgments}
We thank the anonymous referee for helpful comments that improved the clarity of the manuscript. We thank Kevin Aguila for calculating star-formation rates used in this paper. This material is based upon work supported by NASA under an ADAP award (NNH18ZDA001N) and under award number 80GSFC21M0002. The work makes use of data from the Galaxy Evolution Explorer, which was a mission of the National Aeronautics and Space Administration led by the California Institute of Technology. Some of the data presented in this paper were obtained from the Mikulski Archive for Space Telescopes (MAST). This research makes use of the NASA/IPAC Infrared Science Archive and the NASA/IPAC Extragalactic Database, which are funded by the National Aeronautics and Space Administration and operated by the California Institute of Technology. This publication makes use of data products from the Wide-field Infrared Survey Explorer, which is a joint project of the University of California, Los Angeles, and the Jet Propulsion Laboratory/California Institute of Technology, funded by the National Aeronautics and Space Administration. The Infrared Astronomical Satellite (IRAS) was a joint project of the US, UK, and the Netherlands. This research has made use of NASA’s Astrophysics Data System and the HyperLeda database (\url{http://leda.univ-lyon1.fr}).
\end{acknowledgments}

%

\vspace{5mm}
\facilities{\textit{GALEX}, \textit{IRAS}, \textit{WISE}, IRSA}


\software{SExtractor \citep{1996A&AS..117..393B} }




\appendix

\section{Galaxy sample}\label{sec:gal_sample}

\begin{figure*}
\centering
\includegraphics[scale = 0.75]{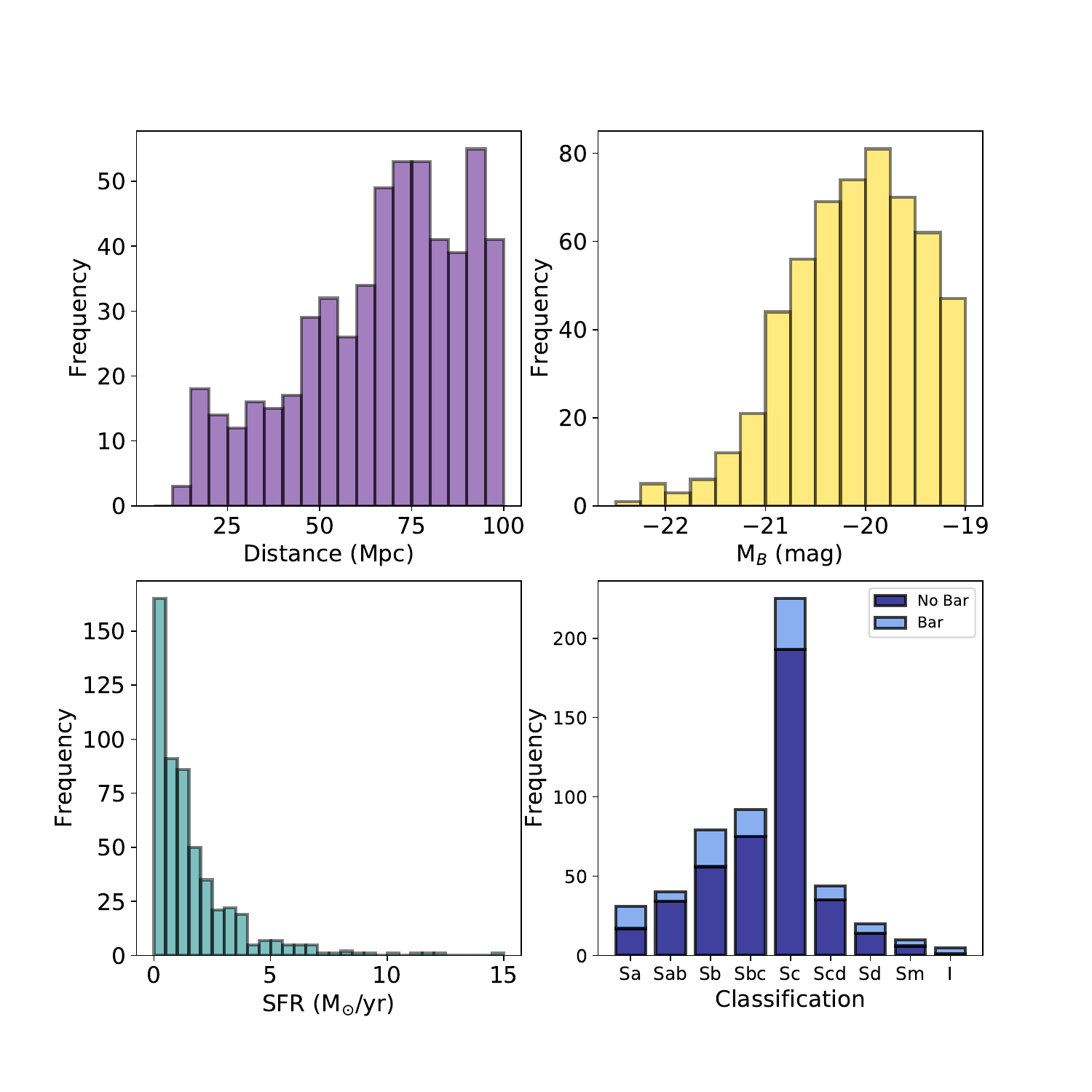}
\caption{Properties of the 551 galaxies in the \textit{GALEX} sample, including their distances (upper left), absolute $B$-band magnitudes (upper right), IR SFRs (lower left), and morphological classifications (lower right). All properties are taken from the HyperLeda database except for the SFRs, which are determined from archival \textit{IRAS} and \textit{WISE} observations as discussed in Appendix \ref{sec:gal_sample}.}
\label{fig:gal_prop}
\end{figure*}

We present the properties of the galaxy sample in Fig.~\ref{fig:gal_prop} and Tab.~\ref{tab:gal_prop}, including the morphological type, right ascension, declination, distance, inclination angle, major and minor axes, absolute $B$-band magnitude, SFR, and \textit{GALEX} FUV and NUV exposure times. We calculate infrared (IR) SFRs for the sample as follows. For the 42/551 galaxies with high- or moderate-quality archival flux densities from the \textit{Infrared Astronomical Satellite} (\textit{IRAS})\footnote{\url{https://irsa.ipac.caltech.edu/IRASdocs/iras.html}}, we adopt SFR[$M_{\odot}$ yr$^{-1}$] $\approx 4.5\times10^{-44}L_{\text{IR}}$[ergs s$^{-1}$] \citep{1998ApJ...498..541K}, where $L_{\text{IR}}$[erg s$^{-1}] \approx 2.2\times10^{39} D$[Mpc]$^{2} (13.48S_{\text{12}} + 5.16S_{\text{25}} + 2.58S_{\text{60}} + S_{\text{100}})$ \citep[e.g.,][]{1996ARA&A..34..749S}. Here, $D$ is the distance in Mpc and $S_{\text{12}}$, $S_{\text{25}}$, $S_{\text{60}}$, and $S_{\text{100}}$ are the $12$~$\mu$m, $25$~$\mu$m, $60$~$\mu$m, and $100$~$\mu$m flux densities measured in Janskys.

\begin{figure*}
\centering
\includegraphics[scale = 0.75]{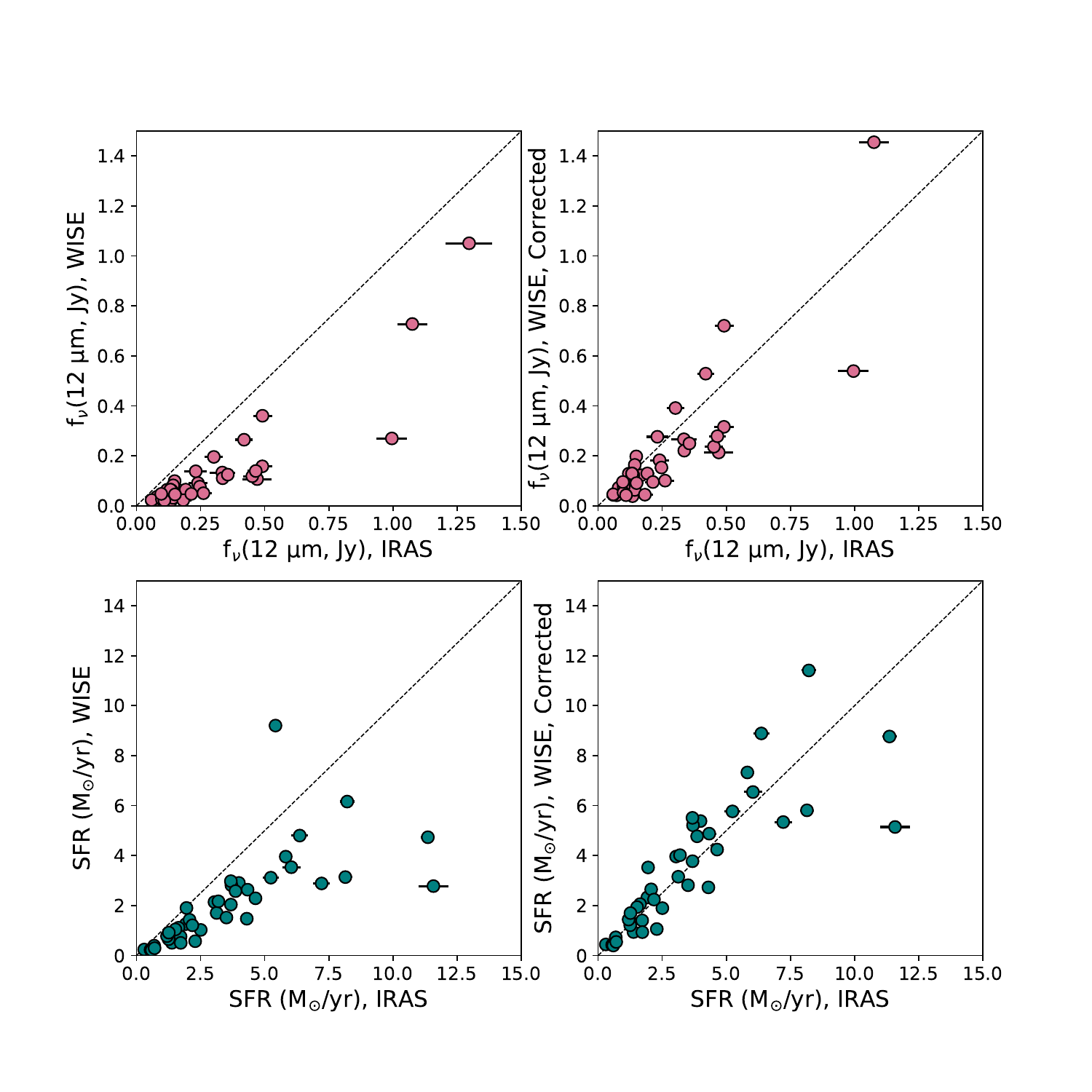}
\caption{Comparison of IR SFRs determined from archival \textit{IRAS} and \textit{WISE} observations of galaxies in the \textit{GALEX} sample. The $12~\mu m$ flux densities and SFRs derived from \textit{WISE} are systematically below those from \textit{IRAS} (upper and lower left panels), which we correct for by applying a scale factor of $\alpha = 2$ to the \textit{WISE} measurements (upper right panel). The SFRs determined from corrected \textit{WISE} flux densities are consistent with those determined from \textit{IRAS} (bottom right panel). The \textit{IRAS} SFRs are calculated following \citet{1998ApJ...498..541K} assuming an $L_{\text{IR}}$ determined from all four \textit{IRAS} bands \citep[e.g.,][]{1996ARA&A..34..749S}, and the \textit{WISE} SFRs are calculated from the $12~\mu m$ flux densities following \citet{2017ApJ...850...68C}. The error bars reflect the measurement error only.}
\label{fig:sfr_calc}
\end{figure*}

For the remaining galaxies in the sample, we calculate a $12$~$\mu$m SFR from archival \textit{Wide-field Infrared Survey Explorer} (\textit{WISE})\footnote{\url{http://wise.ssl.berkeley.edu/mission.html}} data following \citet{2017ApJ...850...68C}. Both the \textit{IRAS} and \textit{WISE} data are retrieved from the NASA/IPAC Infrared Science Archive (IRSA)\footnote{\url{https://irsa.ipac.caltech.edu}}; see \dataset[10.26131/irsa4]{https://doi.org/10.26131/irsa4} and \dataset[10.26131/irsa11]{https://doi.org/10.26131/irsa11} for the \textit{IRAS} data and \dataset[10.26131/irsa1]{https://doi.org/10.26131/irsa1} for the \textit{WISE} data. In Fig.~\ref{fig:sfr_calc}, we compare the $12$~$\mu$m flux densities from \textit{IRAS}, $f_{\nu,12,\text{IRAS}}$, and \textit{WISE}, $f_{\nu,12,\text{WISE}}$, for the 42 galaxies in the sample with data available from both missions. The \textit{WISE} flux densities are systematically below the \textit{IRAS} flux densities by a factor of $\alpha = 2$, which we determine by minimizing
\begin{equation}
\chi^{2} = \sum\limits_{i} \frac{(f_{\nu,12,\text{IRAS},i} - \alpha f_{\nu,12,\text{WISE},i})^{2}}{\sigma_{f_{\nu,12,\text{IRAS}},i}^2}.
\end{equation}

We correct for the difference by applying the scale factor to the \textit{WISE} flux densities before calculating the SFRs, which improves the consistency between the \textit{IRAS} and \textit{WISE} SFR estimates (see Fig.~\ref{fig:sfr_calc}). We find comparable SFRs from the relations determined by \citet{2017ApJ...850...68C} for the W3 $12$~$\mu$m and W4 $23$~$\mu$m bands. There are 15 galaxies in the sample that do not have observations available in either the \textit{IRAS} or \textit{WISE} archives. For these galaxies, we do not report a SFR in Tab.~\ref{tab:gal_prop}, and we do not include them in our SFR stacking analyses.

\begin{deluxetable*}{cccccccccccc}
  \tablecaption{Properties of \textit{GALEX} galaxy sample}
  \tablehead{
    \colhead{(1)} &
    \colhead{(2)} &
    \colhead{(3)} &
    \colhead{(4)} &
    \colhead{(5)} & 
    \colhead{(6)} &
    \colhead{(7)} & 
    \colhead{(8)} &
    \colhead{(9)} &
    \colhead{(10)} &
    \colhead{(11)} &
    \colhead{(12)} \\
    \colhead{Galaxy} &
    \colhead{Type} &
    \colhead{R.A.} &
    \colhead{Decl.} &
    \colhead{$D$} & 
    \colhead{$i$} &
    \colhead{$D_{\text{25}}$} &
    \colhead{$B_{\text{25}}$} &
    \colhead{$M_{\text{B}}$} &
    \colhead{SFR (IR)} &
    \colhead{$t_{\text{exp}}$(NUV)} &
    \colhead{$t_{\text{exp}}$(FUV)}\\
    \colhead{} &
    \colhead{} &
    \colhead{($^{\circ}$; J2000)} &
    \colhead{($^{\circ}$; J2000)} &
    \colhead{(Mpc)} & 
    \colhead{($^{\circ}$)} &
    \colhead{($''$)} & 
    \colhead{($''$)} &
    \colhead{(mag)} &
    \colhead{($M_{\odot}$ yr$^{-1}$)} &
    \colhead{(s)} &
    \colhead{(s)}
  }
  \startdata
  \noalign{\smallskip}
   PGC000012 & Sa & 0.0360 & -6.3739 & 95 & 90 & 84 & 18 & -20.9 & 2.8 & 1551 & 1551 \\ 
   PGC000192 & Sc & 0.7027 & -3.6061 & 89 & 80 & 66 & 18 & -20.7 & 3.2 & 1695 & - \\ 
   UGC00043 & SABd & 1.5101 & 14.4174 & 77 & 90 & 48 & 12 & -19.3 & 0.1 & 1676 & 1676 \\ 
   NGC7831 & Sb & 1.8312 & 32.6092 & 75 & 90 & 105 & 24 & -22.2 & 3.9 & 1688 & - \\ 
   NGC0013 & Sb & 2.1988 & 33.4334 & 72 & 90 & 135 & 36 & -21.1 & 1.1 & 1688 & - \\
   ... & ... & ... & ... & ... & ... & ... & ... & ... & ... & ... & ... \\
  \label{tab:gal_prop}
  \enddata
  \tablecomments{(1) Galaxy name, (2) morphological type, (3) right ascension, (4) declination, (5) distance, (6) inclination angle, (7) major and (8) minor axes, defined by the isophotal level of $\mu_{\text{B}} = 25$ mag arcsec$^{-2}$, (9) absolute $B$-band magnitude, (10) IR SFR, and \textit{GALEX} exposure times in the NUV (11) and FUV (12). The full table for all 551 galaxies in the sample is available in the online-only materials.}
  \tablereferences{Columns (1) - (9) are retrieved from the HyperLeda database (\url{https://leda.univ-lyon1.fr/}), column (10) is calculated from \textit{IRAS} and \textit{WISE} catalogs accessed via IRSA (see Appendix~\ref{sec:gal_sample}; \url{https://irsa.ipac.caltech.edu}), and columns (11) and (12) are determined from the \textit{GALEX} archive (\url{http://galex.stsci.edu/GR6/}).}
\end{deluxetable*}

\section{UV luminosity profiles}\label{sec:lum_prof}

\begin{figure*}
\centering
\includegraphics[scale = 0.75]{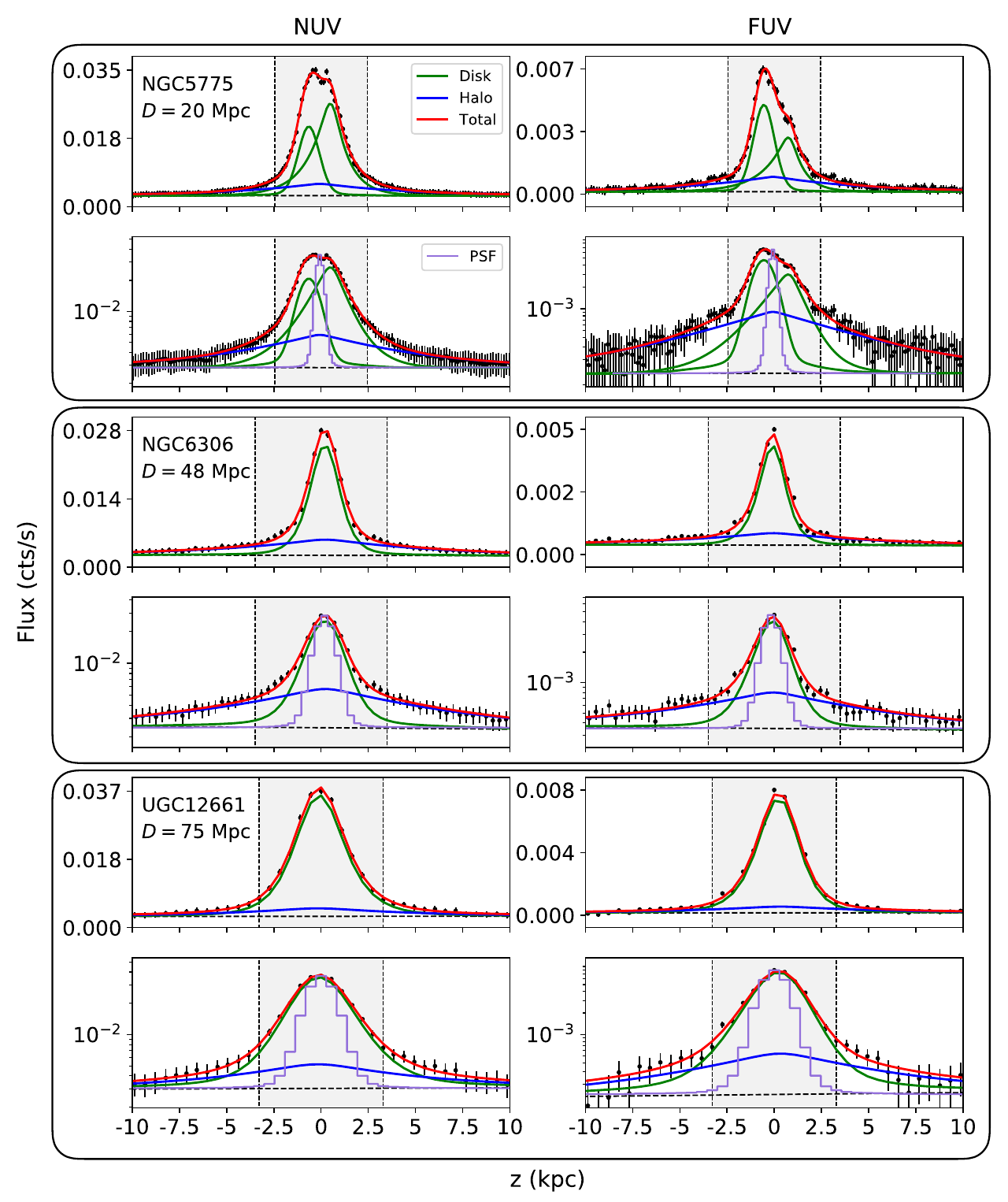}
\caption{Example NUV (left) and FUV (right) profiles of three galaxies with halo detections, displayed on a linear (top) and log scale (bottom). The disk and halo component(s) are shown in green and blue, respectively, and the total flux profile is indicated in red. The dashed black horizontal line marks the background level, and the shaded gray box indicates the region within $|z| < b_{25}$. A cross-section of the \textit{GALEX} PSF is shown by the purple histograms; the detected halos are significantly above the PSF wings.}
\label{fig:gal_profiles}
\end{figure*}

We include supplementary figures showing the vertical luminosity profiles and best-fit models for individual galaxies and galaxy stacks here. In Fig.~\ref{fig:gal_profiles}, we display the FUV and NUV profiles for several galaxies with halo detections. These are selected to show examples of galaxies with both one- and two-component disk models and a range of halo luminosity and S/N.

\begin{figure*}
\centering
\includegraphics[scale = 0.75]{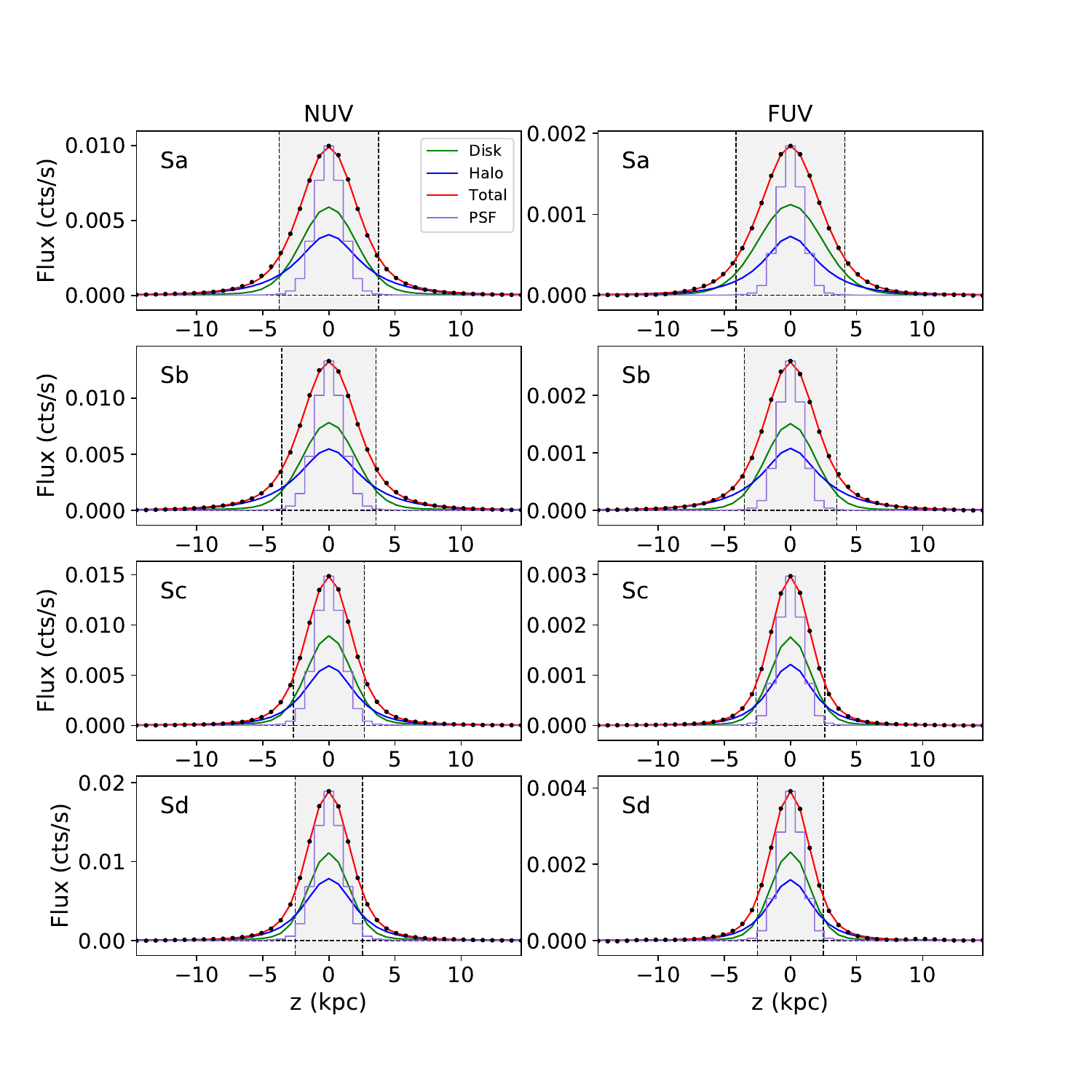}
\caption{Same as Fig.~\ref{fig:stack_Hubble_Sd} for all morphological stacks. UV halos are detected in all morphological types (Sa - Sd, as shown from top to bottom).}
\label{fig:stack_Hubble_all}
\end{figure*}

\begin{figure*}
\centering
\includegraphics[scale = 0.62]{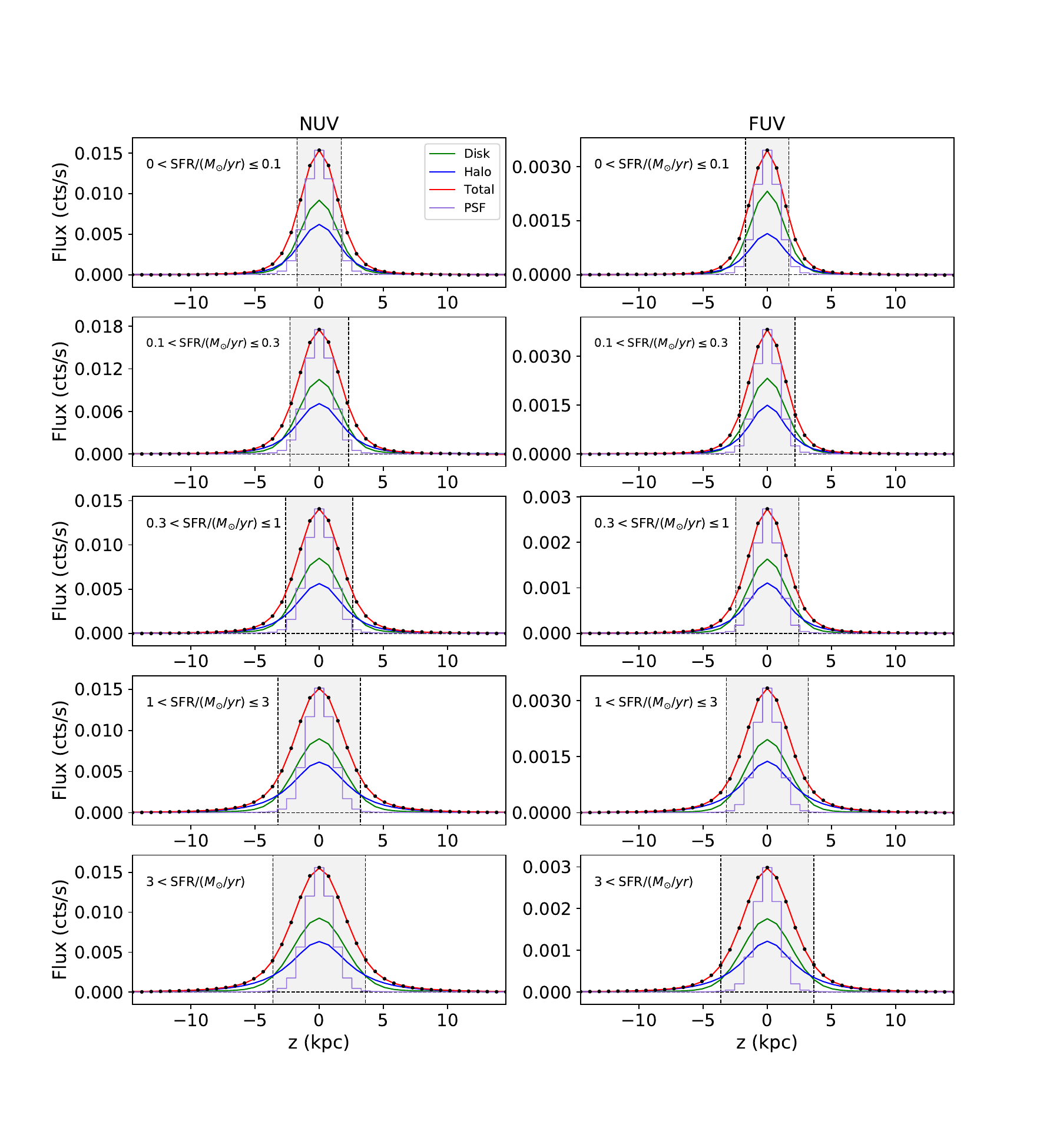}
\caption{Same as Fig.~\ref{fig:stack_Hubble_Sd} for the SFR stacks; UV halos are detected in all SFR stacks.}
\label{fig:stack_SFR_all}
\end{figure*}

The composite profiles and best-fit models for all morphology and SFR stacks are shown in Figs.~\ref{fig:stack_Hubble_all} and \ref{fig:stack_SFR_all}, respectively.

\section{Supplementary FUV results}\label{sec:supp_fuv}

\begin{figure*}
\centering
\includegraphics[scale = 0.75]{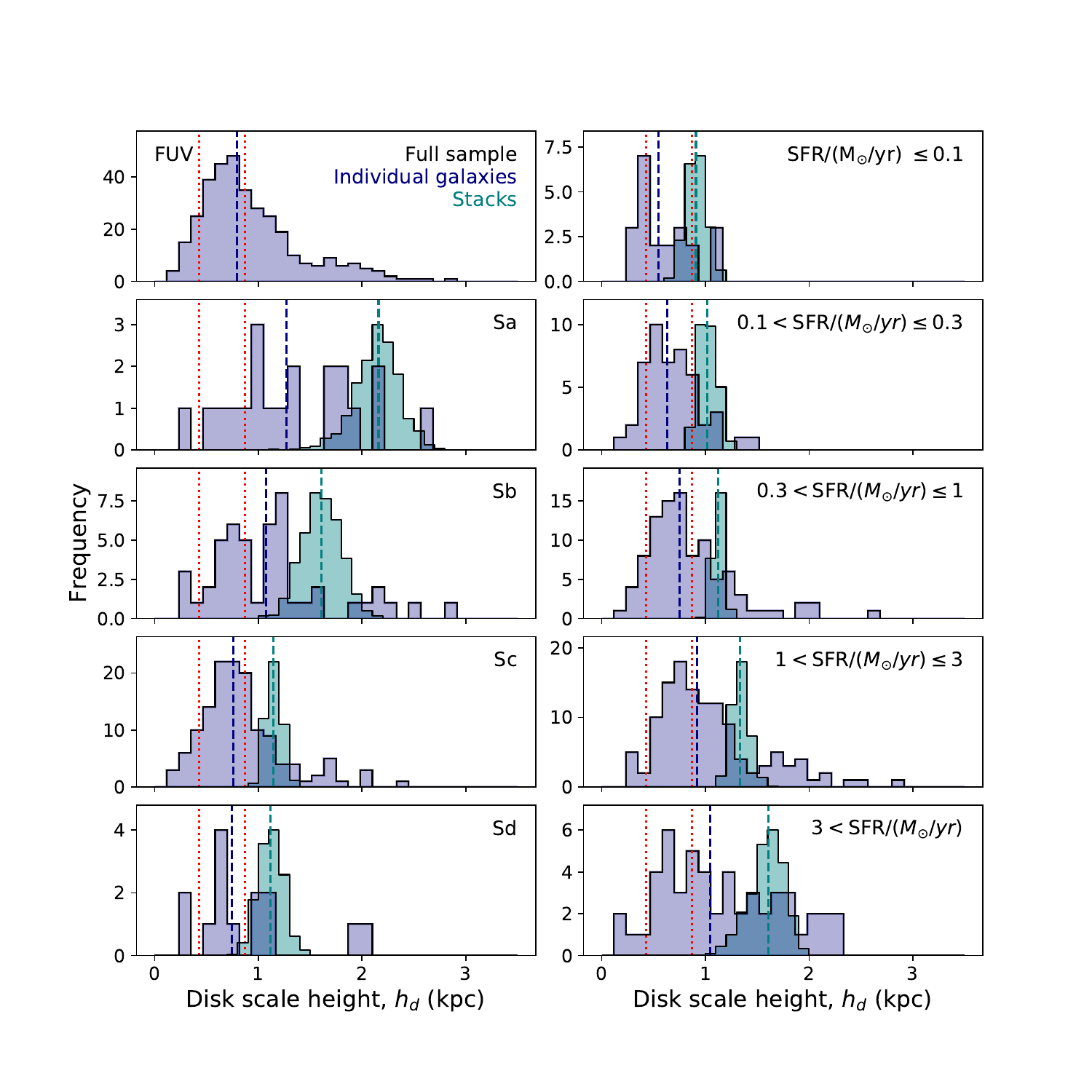}
\caption{Same as Fig.~\ref{fig:hgal_nuv} for the FUV data.}
\label{fig:hgal_fuv}
\end{figure*}

\begin{figure*}
\centering
\includegraphics[scale = 0.75]{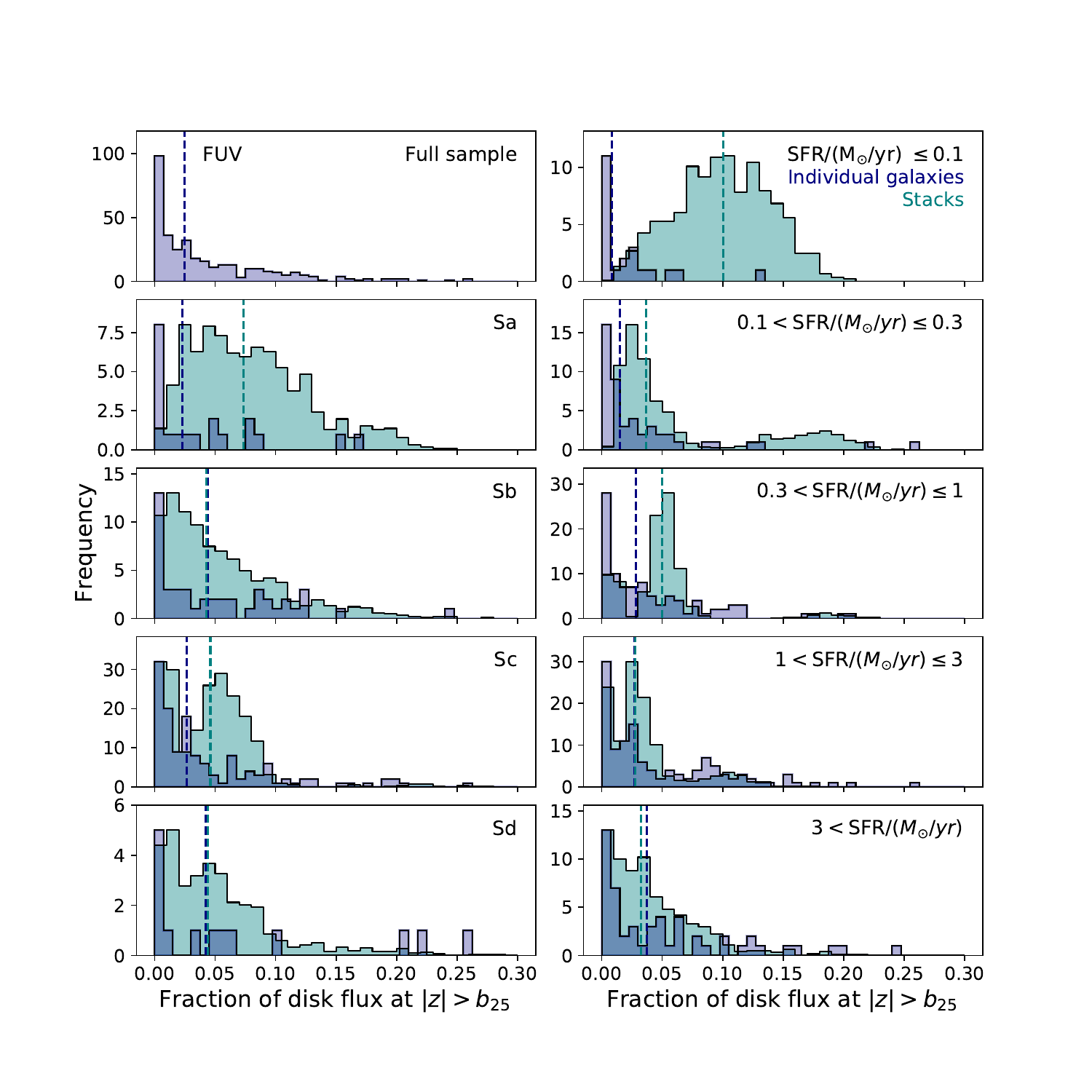}
\caption{Same as Fig.~\ref{fig:f_frac_nuv} for the FUV data.}
\label{fig:f_frac_fuv}
\end{figure*}

We provide equivalent figures for the FUV results as shown above for the NUV here. In Figs.~\ref{fig:hgal_fuv} and \ref{fig:f_frac_fuv}, we show the FUV disk scale heights and fractional disk fluxes at $|z| > b_{\text{25}}$, respectively, for both the individual galaxies and the stacks. 

\section{Extinction Correction}\label{sec:ext_corr}

Here we describe the extinction correction applied to calculate $l_{\text{out}} = L_{\text{h,out}}/L_{\text{in}}$ and constrain the extraplanar dust mass. We perform this correction in the NUV, as approximately half of the halos are detected in galaxies without FUV coverage. This correction is necessary because the extinction of the galaxy light along the major axis ($A_{\text{UV}, \parallel}$, parallel to the disk) is more significant than along the minor axis ($A_{\text{UV}, \perp}$, perpendicular to the disk). Consequently, the extraplanar dust is illuminated by a less attenuated spectrum than we observe along our line of sight.

We follow \citet{1999A&A...352..371B} to estimate $A_{\text{UV}, \parallel}$ from the ratio of the far-IR (FIR) to UV fluxes, $F_{\text{FIR}}/F_{\text{UV}}$ (see their Fig. 1 and Eq. 2). A correlation between $A_{\text{UV}}$ and $F_{\text{FIR}}/F_{\text{UV}}$ is expected because attenuated UV photons are reprocessed into the FIR. There are 147 galaxies in our sample with 60 $\micron$ and 100 $\micron$ flux densities available from IRAS to calculate $F_{\text{FIR}}$ following \citet{1996ARA&A..34..749S}. This yields a median $\langle A_{\text{UV}, \parallel} \rangle = 2.5$ mag for the sample, with a range from $A_{\text{UV}, \parallel} = 1.1 - 5.9$ mag. We then estimate a typical $B$-band extinction along the minor axis of $A_{\text{B}, \perp} \approx 0.2 - 0.5$ mag for Sa - Sd galaxies (see \citealt{2001PASP..113.1449C} and references therein). We expect $A_{\text{UV}}$/$A_{\text{B}} = 2.4$ for Milky Way type dust with $R_{\text{V}} = 3.1$, and therefore we adopt a typical value of $\langle A_{\text{UV}, \perp} \rangle = 1.0$ mag for the sample. We assume that the ratio of the major to minor axis extinction is constant across the sample (i.e., $\langle A_{\text{UV}, \perp} \rangle/\langle A_{\text{UV}, \parallel} \rangle = A_{\text{UV}, \perp}/A_{\text{UV}, \parallel}$ for all galaxies) and thus $A_{\text{UV}, \perp} = A_{\text{UV}, \parallel}/2.5$. The corrected $l_{\text{out}}$ takes the form $l_{\text{out}} = (F_{\text{h,out}}/(F_{\text{tot}} - F_{\text{h,out}}))10^{-0.4(A_{\text{UV}, \parallel} - A_{\text{UV}, \perp})}$. For the median extinction, the correction factor is 0.25, and the characteristic $f_{\text{h,out}}$ becomes $\approx 3$\%. The scatter in the relation of \citet{1999A&A...352..371B} is $\approx 0.3$ mag; accounting for uncertainty in $A_{\text{UV}, \perp}$, the error on the correction factors is generally $\approx 50$\%. 

To estimate an extinction correction for the stacks, we calculate a median $\langle A_{\text{UV}, \parallel} \rangle$ for each morphology and SFR bin based on the available galaxies with IRAS coverage. For the four morphology stacks (Sa - Sd), we find $\langle A_{\text{UV}, \parallel} \rangle = [3.4, 2.6, 2.6, 2.4]$. There is more variation with SFR; from low to high SFR, we find $\langle A_{\text{UV}, \parallel} \rangle = [1.2, 1.2, 1.8, 2.4, 3.3]$ (there are no galaxies with IRAS coverage in the lowest SFR bin, so we adopt the value for the adjacent bin). For individual galaxies without IRAS coverage, we assume the median $\langle A_{\text{UV}, \parallel} \rangle$ based on the galaxy's morphology. There is significant spread in the extinction values in a given bin, with a typical standard deviation of $\approx 1$ mag. This suggests that the correction factors estimated here for galaxies without IRAS coverage are accurate to approximately a factor of two. 

\bibliographystyle{aasjournal}



\end{document}